\documentclass[final,3p,times,11pt]{elsarticle}
\usepackage{natbib}
\usepackage{natbib}
\usepackage{float}   

\usepackage{comment}
\usepackage{amssymb}
\usepackage{svg}
\usepackage{tikz}
\usetikzlibrary{calc}
\usetikzlibrary{shapes.geometric, arrows.meta, positioning}
\usepackage{amsthm}
\usepackage{amsmath}
\usepackage{subcaption}
\usepackage{hyperref}
\usepackage{ mathrsfs }
\usepackage{algorithm}
\usepackage{algpseudocode}
\usepackage{tikz}
\usetikzlibrary{arrows.meta,decorations.pathmorphing}

\journal{Elsevier}

\begin{document}

\begin{frontmatter}



\title{An ensemble-variational approach for open-loop flow control}

 \author[label1,label2,label4]{Riccardo Maranelli}
 \author[label1]{Vincent Mons}
 \author[label3]{Jean-Camille Chassaing}
 \author[label4]{Matthieu Queguineur}
 \author[label2]{Taraneh Sayadi}

 \affiliation[label1]{organization={DAAA, ONERA, Institut Polytechnique de Paris},            
             city={Meudon},
             postcode={92190},            
             country={France}}

 \affiliation[label2]{organization={M2N, CNAM},
             addressline={2 rue Conté},
             city={Paris},
             postcode={75003},            
           country={France}}

 \affiliation[label3]{organization={Sorbonne Université, CNRS, Institut Jean Le Rond d’Alembert},
             addressline={ UMR 7190},
             city={Paris},
             postcode={75005},            
           country={France}}

\affiliation[label4]{organization={Centre National d’Etudes Spatiales CNES},
             addressline={rue Jacques Hillairet},
             city={Paris},
             postcode={75612},             
           country={France}}



\begin{abstract}
The design of effective control strategies for unsteady flows governed by complex, nonlinear dynamics remains a central challenge in fluid mechanics. Adjoint-based optimisation methods, while efficient for high-dimensional problems, require the derivation and implementation of adjoint equations and can exhibit numerical sensitivities in systems that are not smoothly differentiable. This work proposes an ensemble-variational (EnVar) framework as a non-intrusive alternative, in which cost-function gradients are approximated through a finite ensemble of perturbed control vectors, requiring no modification of the forward solver. The present approach is tailored to address high-dimensional control problems, here considering the case where the actuator is represented as a spatially distributed forcing field. The methodology is assessed on two-dimensional open-cavity flows across Reynolds regimes spanning from quasi-periodic to chaotic dynamics. In the quasi-periodic regime, the EnVar framework recovers control strategies consistent with adjoint-based optimisation, achieving significant reductions in kinetic energy fluctuations and driving the flow toward a periodic limit cycle. In the chaotic regime, the framework remains effective in estimating gradient information and mitigating flow fluctuations. Taken together, these results demonstrate that the EnVar method constitutes a computationally efficient, parallelisable, and non-intrusive alternative to adjoint-based approaches for high-dimensional flow optimisation across a broad range of unsteady and chaotic regimes.
\end{abstract}




\begin{keyword}
ensemble variational methods, adjoint, optimisation, flow control
\end{keyword}

\end{frontmatter}


\section{Introduction}
\label{sec:intro}

The prediction and control of unsteady flows with complex dynamics remain central challenges in fluid mechanics, particularly when strong instabilities and nonlinear interactions dominate the system response. Flow instabilities arise in a wide range of engineering systems, degrading performance, reducing efficiency, and compromising operational reliability. Effective mitigation of such instabilities therefore requires robust, systematic control strategies capable of handling the nonlinear and unsteady nature of the underlying flow physics.

In complex flow regimes, the design of effective control strategies typically leads to high-dimensional optimisation problems. Gradient-based methods iteratively minimise a cost function by exploiting descent direction information \cite{Kochenderfer2019}.
When analytical derivatives are available, gradients can be evaluated directly at low computational cost, however, in most practical applications, analytical gradients are not accessible, motivating the use of alternative numerical approaches. A straightforward approach to gradient approximation is the use of finite-difference (FD) methods. While simple to implement, these methods scale poorly with the number of parameters, as each additional design variable requires an extra function evaluation. Similarly, stochastic optimisation methods that estimate gradients approximately \cite{Spall2003} face significant challenges in high-dimensional spaces. Population-based algorithms, such as genetic algorithms and particle swarm optimisation, are capable of exploring complex design spaces; however, they typically require a large number of function evaluations per generation, which makes them computationally prohibitive for high-fidelity CFD simulations. These considerations have led traditional gradient-based optimisation methods in CFD to rely on adjoint solvers \cite{Giles2000}. First introduced for partial differential equations by Lions \cite{Lions1971} and extended to fluid mechanics by Pironneau \cite{Pironneau1974} and Jameson \cite{Jameson1988}, this approach enables sensitivity computation at a cost independent of the number of design variables. Although highly accurate, adjoint methods can be difficult to implement for complex codes, sensitive to numerical errors, and prone to instability in strongly nonlinear or turbulent flows \cite{Pope2000}. In addition, in the case of unsteady problems, the adjoint approach requires the storage, or the recomputation, of the full forward trajectory during the backward-in-time integration of the adjoint problem, thereby restricting their use for long integration times \cite{Vishnampet2015}.

To address the limitations of the aforementioned optimisation approaches while preserving computational tractability and broad applicability, ensemble-based variational (EnVar) methods have recently gained attention. These methods provide a non-intrusive alternative by approximating cost-function gradients through a finite ensemble of perturbed control vectors \cite{Liu2008, Mons2016}. By projecting the optimisation problem onto the subspace spanned by the ensemble, EnVar methods are able to capture nonlinear interactions while keeping the computational cost manageable. Building on the ensemble Kalman methodology introduced by Evensen \cite{Evensen2009}, EnVar techniques have been extended to variational and nonlinear least-squares formulations \cite{Chada2021, Nakano2021, Bannister2017, Ding2021} and have seen increasing use in fluid mechanics. Applications include identifying transition-triggering disturbances in hypersonic boundary layers \cite{Jahanbakhshi2019}, designing optimal thermal actuation for transition delay \cite{Jahanbakhshi2021}, solving inverse problems in transitional flows \cite{Buchta2021}, shape optimisation in unsteady or stochastic environments where adjoint methods are impractical \cite{Lorente2023}, and LES-based optimisation to suppress turbulent wake fluctuations under chaotic dynamics \cite{Zhang2024}. Further developments include passive and active control of vortex shedding \cite{Liu2025}, nonlinear nonmodal optimisation and droplet-impact control \cite{Rodriguez2024}, turbulence-model data assimilation \cite{Mons2021}, and scalar-source identification in turbulent flows \cite{Mons2019}. These studies highlight the broad potential of EnVar approaches for tackling nonlinear optimisation problems in fluid mechanics.


Despite this progress, a systematic EnVar formulation for high-dimensional unsteady flow control, as well as rigorous comparisons with established adjoint-based optimisation methods, remain underdeveloped. Within this context, the present study develops and assesses an ensemble-variational framework, formulated within an ensemble finite-difference setting \cite{bocquet2013}, for flow optimisation in high-dimensional control problems, where the actuator is represented as an (unknown) spatially distributed forcing field rather than through a small set of prescribed actuation parameters.
The optimisation problem to identify such an actuator is made tractable through identifying an appropriate and low-dimensional ensemble subspace. This is achieved by exploiting the Bayesian formulation of the EnVar approach and constructing a physically-motivated prior covariance matrix associated to the actuator.
The ensemble is chosen to span the dominant eigendirections of this covariance matrix, allowing to naturally enforce spatial smoothness in the optimisation procedure, among other desirable properties.

The framework is applied to the open cavity, a canonical separated-flow configuration in which shear-layer instabilities give rise to self-sustained oscillations and, at higher Reynolds numbers, chaotic dynamics. Optimal open-loop control strategies are determined within large control spaces, with the objective of mitigating flow fluctuations across both quasi-periodic and chaotic regimes, thereby assessing the robustness of the approach over a wide range of unsteady behaviours. The performance of the EnVar framework is systematically compared with that of adjoint-based optimisation, providing a rigorous evaluation of its capabilities relative to an established reference method in the context of strongly nonlinear interactions and high-dimensional unsteady flow problems.

This study is organized as follows. Section \textsection~\ref{flow_conf} describes the flow configuration considered in the present study. Section \textsection~\ref{sect3} presents the mathematical formulation of the proposed optimisation framework, detailing both the EnVar approach and the adjoint-based method. Details about the flow-control setup and numerical implementation are provided in Section \textsection~\ref{sec:control_objective_numerics}. The results of the cavity-flow control are discussed in Section \textsection~\ref{sect5}, and concluding remarks are provided in Section \textsection~\ref{sectconcl}.

\section{Flow configuration}\label{flow_conf}


This study focuses on the flow over a two-dimensional open cavity \cite{Sipp2007}, a canonical configuration that illustrates fundamental instability mechanisms and has attracted significant attention for both theoretical and applied research. At sufficiently high Reynolds numbers, the flow is unsteady, exhibiting self-sustained oscillations that correspond to the development of Kelvin–Helmholtz instabilities of the free shear layer spanning the opening in association with a pressure-feedback mechanism \cite{rossiter1962effects}, while recirculating motions develop inside the cavity. The dynamics identify with (quasi-)periodic ones for lower values of the Reynolds number, while transitioning to a chaotic behaviour for larger values. Various control strategies such as model-based feedback, open-loop forcing near Hopf bifurcations, and mean-flow resolvent-based linear models have been successfully applied to stabilize oscillations and suppress quasi-periodic attractors in this configuration  \cite{Sipp2007, Barbagallo2009, Sipp2010, Barbagallo2011, Sipp2012, Leclercq2019}. 
However, the development of control strategies based on numerical models of the nonlinear Navier-Stokes equations, including regimes that may become chaotic, remains an active area of research.

\begin{figure}
    \centering
    \begin{tikzpicture}[scale=2.0,>=Stealth]
        \def\H{0.5}    
        \def\D{1.0}     
        \def\Lone{1.2}  
        \def\Ltwo{1.5}  
        \def\Lones{0.4}  
        \def\Ltwos{0.75}  
        \def\Lcav{1.0}  
        \def\xamin{-0.35}
        \def\xamax{0.15}
        \def\yamin{-0.65}
        \def\yamax{0.2}

        \draw[thick] (-\Lone,0) -- (0,0);
        \draw[thick] (\D,0) -- (\D+\Ltwo,0);
        \draw[thick] (-\Lone,\H) -- (\D+\Ltwo,\H);
        \draw[thick] (-\Lone,0) -- (-\Lone,\H);
        \draw[thick] (\D+\Ltwo,0) -- (\D+\Ltwo,\H);
        \draw[thick] (0,0) -- (0,-\D) -- (\D,-\D) -- (\D,0);

        \fill[gray!50] (-0.015,-\D) rectangle (-0.05,-0.015);
        \fill[gray!50] (\D+0.015,-0.015) rectangle (\D+0.05,-\D); 
        \fill[gray!50] (0,-\D-0.015) rectangle (\D,-\D-0.05);
        \fill[gray!50] (-\Lones,-0.015) rectangle (-0.015,-0.05);
        \fill[gray!50] (0.015+\D,-0.015) rectangle (\D+\Ltwos,-0.05); 

        \draw[<->] (2.2,0) -- (2.2,\H) node[midway,left] {$H$};
        \draw[<->] (-0.2,-\D) -- (-0.2,0) node[midway,left] {$D$};
        \draw[<->] (-\Lone,-0.15) -- (0,-0.15) node[midway,below] {$L_1$};
        \draw[<->] (\D,-0.15) -- (\D+\Ltwo,-0.15) node[midway,below] {$L_2$};

        \draw[<->,gray!130] (-\Lones,-0.15) -- (0,-0.15) node[midway,below,text=gray!130] {$L_{1,\mathrm{slip}}$};
        \draw[<->,gray!130] (\D,-0.15) -- (\D+\Ltwos,-0.15) node[midway,below,text=gray!130] {$L_{2,\mathrm{slip}}$};

        \draw[->,orange,thick] (0.5,-0.2) arc (90:-180:0.3);
        \draw[orange,thick] plot[domain=0.04:0.98,samples=1000]
          ({\x*\D}, {0.01*\x*10*sin(2000*\x)});

        \draw[red,thick,dashed] (\xamin,0) -- (\xamin,\yamax);
        \draw[red,thick,dashed] (\xamin,\yamax) -- (\xamax,\yamax);
        \draw[red,thick,dashed] (\xamax,\yamax) -- (\xamax,\yamin);
        \draw[red,thick,dashed] (\xamax,\yamin) -- (0,\yamin);

        \fill[yellow!80!black] (0.2,0) circle (0.04);
        \fill[yellow!80!black] (0.4,0) circle (0.04);
        \fill[yellow!80!black] (0.6,0) circle (0.04);
        \fill[yellow!80!black] (0.8,0) circle (0.04);

        \node[font=\tiny,color=black] at (-0.1,0.1) {Actuator};
        \foreach \y in {0,0.1,0.2,0.3,0.4,0.5}
          \draw[->,orange,thick] (-\Lone,\y) -- (-\Lone+0.3,\y);

        \node[left,black] at (-\Lone,0.25) {$U$};

        \draw[->,thick] (-1.8,0) -- (-1.5,0) node[below] {$x$};
        \draw[->,thick] (-1.8,0) -- (-1.8,0.3) node[left] {$y$};
        \fill (-1.8,0) circle (0.012);
    \end{tikzpicture}
    \caption{Sketch of the open-cavity flow configuration. The filled yellow circles denote the sensors. The slip portions of the lower wall upstream and downstream of the cavity are indicated by $L_{1,\mathrm{slip}}$ and $L_{2,\mathrm{slip}}$, respectively. The dashed red lines delineate the control region.} 
    \label{fig:cavity_tikz}
\end{figure}
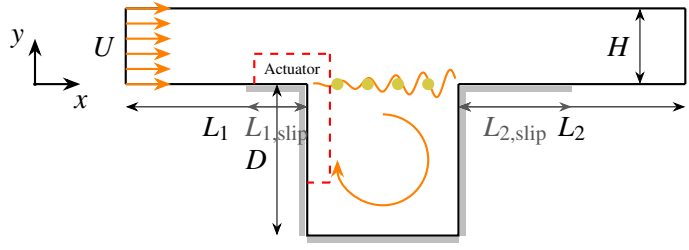

\begin{table}
\centering
\begin{tabular}{lll}
\hline
\textbf{Quantity} & \textbf{Symbol} & \textbf{Value} \\
\hline
Upstream cavity length & $L_1$ & $1.2$ \\
Upstream slip length & $L_{1,\mathrm{slip}}$ & $0.4$ \\
Downstream cavity length & $L_2$ & $1.5$ \\
Downstream slip length & $L_{2,\mathrm{slip}}$ & $0.75$ \\
Cavity depth & $D$ & $1$ \\
Cavity height & $H$ & $0.5$ \\
\hline
\end{tabular}
\caption{Geometrical dimensions of the cavity \cite{Sipp2007}.}
\label{tabgeometry}
\end{table}

A sketch of the present square open cavity configuration is shown in Figure~\ref{fig:cavity_tikz}. Its dimensions are reported in Table \ref{tabgeometry} and correspond to the same ones as in \cite{Sipp2007}. Boundary conditions are specified in Section \ref{sec:BC_numerical_methods}. The reference scales are the uniform velocity \(U\) at the inlet, the length \(D\) of the square cavity, and the kinematic viscosity \(\nu\), leading to the definition of the Reynolds number $Re=U D/\nu$. The underlying flow is governed by the two-dimensional incompressible forced Navier--Stokes equations, written below in non-dimensional form
 \begin{equation}\label{eq:forced_NS}
\begin{aligned}
\begin{cases}
\dfrac{\partial\boldsymbol{u}}{\partial t} +(\boldsymbol{u} \cdot \nabla) \boldsymbol{u}+ \nabla p - \dfrac{1}{Re} \Delta \boldsymbol{u}=\boldsymbol{f}_{\boldsymbol{u}}, \\
    \nabla \cdot \boldsymbol{u}=0,
\end{cases}
\end{aligned}
\end{equation}
where $\boldsymbol{u}$ is the two-dimensional velocity vector, $p$ is the pressure, and $\boldsymbol{f}_{\boldsymbol{u}}$ represents an external force acting on the flow. This forcing $\boldsymbol{f}_{\boldsymbol{u}}=\boldsymbol{f}_{\boldsymbol{u}}(\boldsymbol{x})$ corresponds to a spatially-distributed, steady field. It represents the actuation mechanism from a flow control perspective, which, as further detailed in Section \ref{controlstrategy}, is restricted to a region around the leading edge of the cavity, as depicted by the dashed red lines in Figure \ref{fig:cavity_tikz}, to effectively influence the downstream flow development.

The next section is dedicated to the presentation of methodologies to determine the forcing field $\boldsymbol{f}_{\boldsymbol{u}}$ in order to reach a flow-control objective. These methodologies will be more conveniently derived relying on a semi-discrete formulation where the forced Navier-Stokes equations \eqref{eq:forced_NS} are spatially discretised. The corresponding governing equations may be written in the following compact form
\begin{equation}\label{eq:discrete_NS}
\mathbf{L}\frac{\partial\boldsymbol{q}}{\partial t}
+
\boldsymbol{N}(\boldsymbol{q})
=
\mathbf{P}\boldsymbol{f},
\end{equation}
where the vector $\boldsymbol{q}$ gathers the degrees of freedom for the velocity and pressure variables, while $\boldsymbol{f}$ gathers those for the forcing $\boldsymbol{f}_{\boldsymbol{u}}$. Matrix $\mathbf{L}$ restricts the application of the time derivative to the momentum equations, matrix $\mathbf{P}$ makes the forcing act on the momentum equations only, while $\boldsymbol{N}(\boldsymbol{q})$ corresponds to the discrete counterpart of all remaining contributions.

\section{Optimisation methodologies for open-loop flow control}\label{sect3}

This section presents the mathematical formulation of the objective function used in the optimisation procedure, followed by a description of the adjoint-based and ensemble variational approaches employed for open-loop flow control. 

\subsection{Bayesian formulation of the flow-control problem}

We seek to determine an optimal forcing $\boldsymbol{f}$ appearing in the semi-discrete Navier-Stokes equations \eqref{eq:discrete_NS} such that prescribed target values $\boldsymbol{y}$ are achieved for selected quantities of interest, ranging from the full flow field to integrated global quantities. In this study, $\boldsymbol{y}$ corresponds to target values for velocity fluctuations at some pointwise locations, as fully specified in Section \ref{controlstrategy}. Such an inverse problem may be formulated within a Bayesian formalism considering both $\boldsymbol{f}$ and $\boldsymbol{y}$ as random variables. Within this framework, determining the control strategy amounts to characterizing the posterior distribution $\pi(\boldsymbol{f}|\boldsymbol{y})$, i.e., the probability of the forcing $\boldsymbol{f}$ conditioned on the target $\boldsymbol{y}$. Bayes' theorem relates this posterior to the likelihood $\pi(\boldsymbol{y}|\boldsymbol{f})$, representing the probability that the target $\boldsymbol{y}$ is achieved under the actuation $\boldsymbol{f}$, and the prior distribution $\pi(\boldsymbol{f})$, which encodes any desirable properties to be enforced on the retrieved forcing. This reads as
\begin{equation}
    \pi(\boldsymbol{f}|\boldsymbol{y})\propto \pi(\boldsymbol{y}|\boldsymbol{f}) \pi(\boldsymbol{f}),
\end{equation}
where the proportionality coefficient corresponds to the inverse of the distribution $\pi(\boldsymbol{y})=\int p(\pi(\boldsymbol{y}|\boldsymbol{f}) \pi(\boldsymbol{f})d\boldsymbol{f}$, and can be considered only as a normalization constant. Assuming that both $\pi(\boldsymbol{y}|\boldsymbol{f})$ and $\pi(\boldsymbol{f})$ correspond to Gaussian distributions, the posterior distribution may be written as
\begin{equation}\label{eq:gaussian_posterior}
    \pi(\boldsymbol{f}|\boldsymbol{y})\propto \exp(-J(\boldsymbol{f})),
\end{equation}
where the cost function $J(\boldsymbol{f})$ is given by
\begin{equation}\label{eq:Envarcost}
J(\boldsymbol{f}) = \frac{1}{2} \left\| \boldsymbol{y} - \mathbf{H}\boldsymbol{q}(\boldsymbol{f})\right\|_{\mathbf{R}^{-1}}^2 
                + \frac{1}{2} \left\| \boldsymbol{f} - \boldsymbol{f}^{(e)} \right\|_{\mathbf{B}^{-1}}^2.
\end{equation}
The first contribution on the right-hand-side of \eqref{eq:Envarcost} originates from the likelihood $\pi(\boldsymbol{y}|\boldsymbol{f})$ and quantifies the discrepancies between the target $\boldsymbol{y}$ and the analogous quantities of interest for the controlled flow with actuation $\boldsymbol{f}$. The operator $\mathbf{H}$ thus maps the full flow state $\boldsymbol{q}$ to such quantities of interest. This operator, considered linear, will be referred to as the observation operator in the following. The notation $\boldsymbol{q}(\boldsymbol{f})$ in \eqref{eq:Envarcost} emphasizes the dependency of the flow state to the forcing $\boldsymbol{f}$. This contribution also involves the covariance matrix $\mathbf{R}$, noting that $\left\| \circ \right\|_{\mathbf{R}^{-1}}^2=\circ^{\mathrm{T}}\mathbf{R}^{-1}\circ$, and reflects the confidence about the target $\boldsymbol{y}$. The second contribution in the right-hand-side of \eqref{eq:Envarcost} arises from the prior $\pi(\boldsymbol{f})$. It involves the vector $\boldsymbol{f}^{(e)}$, which may be interpreted as a first-guess value for the forcing $\boldsymbol{f}$, and the prior covariance matrix $\mathbf{B}$. This matrix allows us to control the intensity and the spatial smoothness of the retrieved forcing. 

It may be noted that the above Bayesian formalism corresponds to a more general statement of the presented inverse problem as opposed to a deterministic formulation, which can be recovered from $\eqref{eq:Envarcost}$ by removing the covariance matrices $\mathbf{R}$ and $\mathbf{B}$. A key advantage of the Bayesian formulation is the natural introduction of the prior covariance matrix $\mathbf{B}$, which provides a principled means of encoding more physical constraints on the forcing $\boldsymbol{f}$ beyond what standard penalization terms can offer, and plays an important role in the ensemble-based approach discussed in the following. In addition, the Bayesian formalism enables uncertainty quantification for the retrieved forcing $\boldsymbol{f}$, as detailed below.  

It follows from \eqref{eq:gaussian_posterior} that the optimal actuation $\boldsymbol{f}$, defined as the maximizer of the posterior density $\pi(\boldsymbol{f}|\boldsymbol{y})$, i.e., the maximum a posteriori (MAP) estimate, coincides with the minimiser of the cost function $J(\boldsymbol{f})$. The remainder of this section is devoted to discussing optimisation strategies for performing this minimisation.

\subsection{Adjoint-based approach}\label{sec:adjoint_approach}

The first approach considered here to perform the minimisation of the cost function $J(\boldsymbol{f})$ in \eqref{eq:Envarcost} relies on the adjoint technique, see, e.g., \cite{Costanzo2022}. Such an approach may be considered as a relatively standard one to solve optimisation problems for fluid mechanics applications, especially when dealing with high-dimensional problems, keeping in mind that $\boldsymbol{f}$ refers to the discrete counterpart of a spatially-varying actuator. The adjoint technique will be treated as a reference for the ensemble-based variational approach discussed below.

Minimising the cost function $J(\boldsymbol{f})$ under the constraint of the forced Navier-Stokes equations \eqref{eq:discrete_NS} leads to the introduction of the following Lagrangian 
\begin{equation}
\mathscr{L}
=
J
- \int_0^T
\left(
\mathbf{L}\frac{\partial \boldsymbol{q}}{\partial t}
+
\boldsymbol{N}(\boldsymbol{q})
-
\mathbf{P}\boldsymbol{f}
\right)^{\mathrm{T}} \boldsymbol{q}^\dagger \,dt,
\end{equation}
where $T$ corresponds to the time horizon over which the optimisation is performed, and $\boldsymbol{q}^\dagger$ is referred to as the adjoint state. Differentiating the Lagrangian $\mathscr{L}$ with respect to the state vector $\boldsymbol{q}$ yields the following governing equation for the adjoint state
\begin{equation}
-\mathbf{L}^{\mathrm{T}}
\frac{\partial \boldsymbol{q}^{\dagger}}{\partial t}
+
\left(\frac{\partial \boldsymbol{N}}{\partial \boldsymbol{q}}\bigg|_{\boldsymbol{q}}\right)^{\mathrm{T}}
\boldsymbol{q}^{\dagger}
= \mathbf{M}_q
\mathbf{H}^{\dagger}(\mathbf{H}\boldsymbol{q}-\boldsymbol{y}),
\label{eq:adjoint_discrete}
\end{equation}
where the matrix $\mathbf{M}_q$ defines the scalar product associated with the norm $\left\| \circ \right\|_{\mathbf{M}_q}^2=\circ^{\mathrm{T}}\mathbf{M}_q\circ$, which depends on the spatial discretisation approach considered. The adjoint problem \eqref{eq:adjoint_discrete} also involves the adjoint operator $\mathbf{H}^{\dagger}$ which is associated with the observation operator and satisfies $\boldsymbol{z}^{\mathrm{T}}\mathbf{R}^{-1}\mathbf{H}\boldsymbol{p}=\int_{0}^{T}\boldsymbol{p}^{\mathrm{T}}\mathbf{M}_q  \mathbf{H}^{\dagger}\boldsymbol{z}\, dt $ for all admissible target $\boldsymbol{z}$ and state $\boldsymbol{p}$. The adjoint problem is solved backward in time, from \(t=T\) to \(t=0\), and requires the state $\boldsymbol{q}$ to evaluate the right-hand-side of \eqref{eq:adjoint_discrete} and the contribution $((\partial \boldsymbol{N}/\partial \boldsymbol{q})|_{\boldsymbol{q}})^{\mathrm{T}}$, which arises from the linearisation of the model operator $\boldsymbol{N}$.

Once the adjoint state $\boldsymbol{q}^{\dagger}$ is available, the gradient of the Lagrangian with respect to the forcing $\boldsymbol{f}$ can be evaluated according to
\begin{equation}\label{eq:gradient_forcing}
\frac{\partial \mathscr{L}}{\partial \boldsymbol{f}}
=
\mathbf{M}_f^{-1}\left(\int_{0}^{T}\mathbf{P}^{\mathrm{T}}\boldsymbol{q}^\dagger\, dt
+
\mathbf{B}^{-1}\left(\boldsymbol{f}-\boldsymbol{f}^{(e)}\right)\right),
\end{equation}
where the matrix $\mathbf{M}_f$ is associated to the scalar product of the forcing $\boldsymbol{f}$ in a similar way as the matrix $\mathbf{M}_q$ for the flow state.

The above expressions are exploited in an iterative gradient-based minimisation procedure as follows. Starting from the first guess $\boldsymbol{f}^{(e)}$, each iteration of the minimisation procedure first requires solving the Navier-Stokes equations \eqref{eq:discrete_NS} for the current estimation of the forcing $\boldsymbol{f}$. This enables the backward integration in time of the adjoint governing equations \eqref{eq:adjoint_discrete}. The gradient \eqref{eq:gradient_forcing} can then be computed and used to update the estimation of $\boldsymbol{f}$, which is here achieved relying on the low-memory Broyden-Fletcher-Goldfarb-Shanno (L-BFGS) algorithm \cite{Liu1989}. The procedure is then repeated until convergence.

\subsection{Ensemble-based variational approach}

\subsubsection{EnVar scheme}\label{sec:EnVar_scheme}

As an alternative to the above adjoint approach to identify an optimal actuation $\boldsymbol{f}$, we here consider the use of ensemble-based variational (EnVar) techniques \cite{Liu2008,Mons2021}. The EnVar method used here is strongly inspired by \cite{bocquet2013}.

The starting point of the EnVar approach is to look for an optimal forcing $\boldsymbol{f}$ in a subspace that is, ideally, of low dimension and spanned by the so-called ensemble members, which correspond to realizations of $\boldsymbol{f}$ that should be representative of the prior statistics. Namely, we look for an optimal forcing of the form
\begin{equation}\label{eq:ensemble_representation}
\boldsymbol{f}=\boldsymbol{f}^{(e)}+\mathbf{E}\boldsymbol{w},
\end{equation}
where the columns of $\mathbf{E}$ are formed by the $N_{ens}$ ensemble members and the vector $\boldsymbol{w}$ corresponds to coefficients in this ensemble basis. The construction of the ensemble basis $\mathbf{E}$ as used in this study will be detailed in Section \ref{sec:ensemblematrix}. $\boldsymbol{w}$ thus becomes the actual control vector in the EnVar approach. The fact that the ensemble members should be representative of the prior statistics may be expressed as
\begin{equation}\label{eq:ensemble_covariance}
\begin{aligned}
\frac{1}{N_{ens}-1} \, \mathbf{E} \, \mathbf{E}^\mathrm{T} \simeq \mathbf{B},
\end{aligned}
\end{equation}
where the factor $1/(N_{ens}-1)$ could actually be absorbed here in  the definition of $\mathbf{E}$ but is kept nonetheless to make the formulation of the present EnVar scheme closer to that in previous studies.

Based on \eqref{eq:ensemble_representation}, the cost function $J$ in \eqref{eq:Envarcost} may be rewritten according to
 \begin{equation}\label{eq:J_w}
     J(\boldsymbol{w})=\frac{1}{2} \left\| \boldsymbol{y} - \mathbf{H}\boldsymbol{q}(\boldsymbol{f}^{(e)}+\mathbf{E}\boldsymbol{w}) \right\|_{\mathbf{R}^{-1}}^2+\frac{1}{2}(N_{ens}-1)\boldsymbol{w}^{\mathrm{T}}\boldsymbol{w},
 \end{equation}
which we aim to minimise with respect to $\boldsymbol{w}$. The gradient of $J$ with respect to $\boldsymbol{w}$ is given by 
 \begin{equation}\label{eq:gradient_EnVar}
 \frac{\partial J}{\partial \boldsymbol{w}}
=
\mathbf{Y}^{\mathrm{T}} \mathbf{R}^{-1}
\left(
\mathbf{H}\boldsymbol{q}(\boldsymbol{f}^{(e)} + \mathbf{E}\boldsymbol{w})
-
\boldsymbol{y}
\right)
+
(N_{ens}-1)\,\boldsymbol{w}, 
\end{equation}
where the matrix $\mathbf{Y}$ is formally given by
\begin{equation}
    \mathbf{Y}=\mathbf{H}\frac{\partial \boldsymbol{q}}{\partial \boldsymbol{f}}\mathbf{E}.
\end{equation}
The columns of the matrix $\mathbf{Y}$ thus correspond to directional derivatives of the estimated quantities of interest in the directions given by the ensemble members. Instead of relying on the adjoint approach to bypass the evaluation of the sensitivity $\partial \boldsymbol{q}/\partial \boldsymbol{f}$, the matrix $\mathbf{Y}$ is here evaluated based on finite differences. Namely, the $i$th column of $\mathbf{Y}$, which is denoted as $\mathbf{Y}_{(:, i)}$, is computed according to 
\begin{equation}
\begin{aligned}
\mathbf{Y}_{(:, i)}
\simeq
\frac{
\mathbf{H}\boldsymbol{q}(
\boldsymbol{f}^{(e)} + \mathbf{E}\boldsymbol{w} 
+ \varepsilon \mathbf{E}_{(:, i)}
)
-
\mathbf{H}\boldsymbol{q}( 
\boldsymbol{f}^{(e)} + \mathbf{E}\boldsymbol{w} 
)
}{\varepsilon}, 
\end{aligned}
\label{eq:fdapprox}
\end{equation}
where $\varepsilon$ is a small positive real parameter. Assembling the matrix $\mathbf{Y}$ constitutes the most expensive step of the EnVar approach, as it requires performing one CFD calculation to evaluate the estimated quantities of interest for the current estimation of the forcing $\boldsymbol{f}$, and $N_{ens}$ supplementary calculations to compute the same quantities for perturbed forcings in directions given by the ensemble members (columns of $\mathbf{E}$). The evaluation of the matrix $\mathbf{Y}$ also gives access to the Gauss-Newton approximation of the Hessian of the cost function $J(\boldsymbol{w})$, i.e. where contributions arising from the nonlinearities in the relationship between the forcing $\boldsymbol{f}$ and the estimated quantities of interest are neglected, according to
\begin{equation}\label{eq:Hessian_EnVar}
\begin{aligned}
\boldsymbol{\mathcal{H}}_{\mathrm{GN}}
&=
\mathbf{Y}^{\mathrm{T}}\, \mathbf{R}^{-1}\, \mathbf{Y}
+
(N_{ens}-1)\,\mathbf{I}.
\end{aligned}
\end{equation}
It may be noted that, at the end of the optimisation, this Hessian matrix may be used to evaluate the covariance matrix that is associated to the optimal forcing $\boldsymbol{f}$ according to 
\begin{equation}\label{eq:posterior_covariance_matrix}
    \mathbf{B}_{post}=\mathbf{E}\boldsymbol{\mathcal{H}}^{-1}_{\mathrm{GN}}\mathbf{E}^{\mathrm{T}}.
\end{equation}
The gradient \eqref{eq:gradient_EnVar} and Hessian matrix \eqref{eq:Hessian_EnVar} are employed to perform the minimisation of the cost function $J(\boldsymbol{w})$. Similar to the adjoint method in Section \ref{sec:adjoint_approach}, we rely on an iterative gradient-based approach where, at each iteration $j$ of the optimisation procedure, the weights $\boldsymbol{w}$ in the ensemble basis are updated according to
\begin{equation}
\begin{aligned}
\boldsymbol{w}^{[j+1]} &= \boldsymbol{w}^{[j]} + \delta \boldsymbol{w}^{[j]},
\end{aligned}
\end{equation}
where $\delta \boldsymbol{w}^{[j]}$ refers to the update in the weights. Two approaches are considered here to compute the update $\delta \boldsymbol{w}^{[j]}$. The first one, as considered in \cite{bocquet2013}, consists in exploiting the Gauss-Newton approximated Hessian in \eqref{eq:Hessian_EnVar} as
\begin{equation}\label{eq:update_Newton}
\begin{aligned}
\delta \boldsymbol{w}^{[j]}
&= -  \boldsymbol{\mathcal{H}}_{\mathrm{GN}}^{-1}
\frac{\partial J}{\partial \boldsymbol{w}},
\end{aligned}
\end{equation}
where, implicitly, the involved gradient and Hessian are evaluated at $\boldsymbol{w}=\boldsymbol{w}^{[j]}$. The second approach relies on the L-BFGS algorithm, as for the adjoint method, which, for the EnVar approach, amounts to 
\begin{equation}\label{eq:update_BFGS}
\delta \boldsymbol{w}^{[j]} = - \alpha^{[j]} \boldsymbol{\mathcal{H}}_{\mathrm{BFGS}}^{-1}\frac{\partial J}{\partial \boldsymbol{w}},
\end{equation}
where $\boldsymbol{\mathcal{H}}_{\mathrm{BFGS}}$ refers to the Hessian as approximated in the L-BFGS approach, and $\alpha^{[j]}$ is a step size that is determined through a line-search algorithm \cite{Armijo1966_pjm}. The Newton update \eqref{eq:update_Newton} is often used in ensemble approaches, while the use of the L-BFGS algorithm \eqref{eq:update_BFGS} is expected to provide greater robustness in the case where the cost function $J(\boldsymbol{w})$ is far from being quadratic. The benefit in considering this latter update over the Newton one for the present flow-control problem will be investigated in the following.

\subsubsection{Construction of the ensemble matrix}\label{sec:ensemblematrix}

The design of the covariance matrix $\mathbf{B}$ and the construction of the ensemble matrix $\mathbf{E}$ from $\mathbf{B}$ are crucial aspects of the EnVar approach. In the present case where the control vector $\boldsymbol{f}$ corresponds to the discrete counterpart of a spatially-dependent field, the specification of the covariance matrix $\mathbf{B}$ allows us to enforce desirable properties for the optimal actuation, such as spatial smoothness, while also constraining its amplitude. For the present two-dimensional flow configuration, a common choice for the components of  $\mathbf{B}$ may correspond to the following Gaussian kernel 
\begin{equation}
\begin{aligned}
B_{ij}=\sqrt{\sigma_i^2 \sigma_j^2} \exp\left({-\frac{(x_i-x_j)^2}{a_x^2}} -\frac{(y_i-y_j)^2}{a_y^2} -\frac{2(x_i-x_j)(y_i-y_j)}{c_{xy}}\right),
\end{aligned}
\label{eq:covarianceB}
\end{equation}
where $x_i$ and $y_i$ refer to the coordinates of the $i$th mesh point in the discretisation of the forcing $\boldsymbol{f}_{\boldsymbol{u}}(\boldsymbol{x})$. In addition, $\sigma_i^2$ is a variance parameter that allows us to adjust the degree of penalization of each degree of freedom $i$ with the contribution of the prior term to the cost function $J$ scaling as the inverse of this parameter. The parameters $a_x$ and $a_y$ control the spatial correlation lengths in the two coordinate directions, and the parameter $c_{xy}$ allows us to adjust the cross-correlation between them. While the Gaussian kernel \eqref{eq:covarianceB} is adopted throughout this study, this choice is not unique, and alternative kernels, such as those from the Matérn family, to which the Gaussian kernel belongs, may equally be considered, potentially alongside scalable implementations as discussed in \cite{Lindgren2011_jrss}. The choice of the various parameters in \eqref{eq:covarianceB} will be discussed in Section \ref{sec:prior_parameters}.

Once the covariance matrix $\mathbf{B}$ has been specified, one can proceed with the construction of the ensemble matrix $\mathbf{E}$, which should be representative of the prescribed prior information following \eqref{eq:ensemble_covariance}. Most ensemble-based approaches construct the columns of $\mathbf{E}$ as random draws from a Gaussian distribution with zero mean and covariance matrix $\mathbf{B}$, representative of the quantity to be inferred, here the forcing $\boldsymbol{f}$. Alternatively, the present study adopts a deterministic approach where the ensemble members are chosen as the principal directions of $\mathbf{B}$. Namely, the following eigendecomposition is first performed
\begin{equation}
    \mathbf{B}=\mathbf{U}\mathbf{\Sigma}\mathbf{U}^{\mathrm{T}}, \quad \mathbf{U}^{\mathrm{T}}\mathbf{M}_f\mathbf{U}=\mathbf{I},
\end{equation}
where the second equality follows from the orthonormality of the eigenvector matrix $\mathbf{U}$ with respect to the scalar product induced by the matrix $\mathbf{M}_f$. The diagonal matrix $\mathbf{\Sigma}$ is formed by the eigenvalues of $\mathbf{B}$ and is implicitly stored in decreasing order. The ensemble matrix $\mathbf{E}$ may then be built considering the first $N_{ens}$ dominant eigenvectors according to
\begin{equation}\label{eigenvm1}
\begin{aligned}
\mathbf{E} &= \mathbf{U}_{N_{ens}} \, \mathbf{\Sigma}_{N_{ens}}^{1/2} \, \sqrt{N_{ens}-1}.
\end{aligned}
\end{equation}
Here, $\mathbf{U}_{N_{ens}}$ is the first $N_{ens}$ columns of $\mathbf{U}$ and $\mathbf{\Sigma}_{N_{ens}}$ is a diagonal matrix containing the corresponding eigenvalues. This approach therefore ensures that, for a given ensemble size $N_{ens}$, the different ensemble members (columns of $\mathbf{E}$) are orthogonal and optimally span the subspace of admissible forcings $\boldsymbol{f}$ as defined through the covariance matrix $\mathbf{B}$.

\begin{figure}
\centering
\begin{tikzpicture}[
node distance=16mm and 20mm, 
box/.style={rectangle, draw=black, rounded corners, minimum width=35mm, minimum height=9mm, align=center, fill=white},
smallbox/.style={rectangle, draw=black, rounded corners, minimum width=28mm, minimum height=8mm, align=center, fill=white, font=\footnotesize},
  arr/.style={-{Stealth[length=3mm]}, very thick},
  lab/.style={font=\footnotesize, midway, below}
]

\node[box] (ens) {Ensemble generation: \\ $\mathbf{U}, \mathbf{\Sigma}$  from covariance $\mathbf{B}$\\$\mathbf{E} = \mathbf{U}_{N_{ens}} \, \mathbf{\Sigma}_{N_{ens}}^{1/2} \, \sqrt{N_{ens}-1},$ \\ $\boldsymbol{w}=\boldsymbol{0},  \quad \boldsymbol{f}=\boldsymbol{f}^{(e)}+\mathbf{E}\boldsymbol{w}$};
\node[box, right=12mm of ens] (prop) {
$N_{\mathrm{ens}}+1$ forward evaluations: \\[1mm]
Mean evaluation:
$\mathbf{H}\boldsymbol{q}\!\left(
\boldsymbol{f}^{(e)}+\mathbf{E}\boldsymbol{w}^{[j]}
\right)$ \\[1mm]
$N_{\mathrm{ens}}$ evaluations, for $i=1,\ldots,N_{\mathrm{ens}}$: \\
$\mathbf{H}\boldsymbol{q}\!\left(
\boldsymbol{f}^{(e)}+\mathbf{E}\boldsymbol{w}^{[j]}
+\varepsilon \mathbf{E}_{(:,i)}
\right)$
};
\node[smallbox, right=12mm of prop] (cov) {Compute gradient $\dfrac{\partial J}{\partial \boldsymbol{w}}$  \\  (and Hessian $\mathcal{H}$)};

\node[box, below=18mm of cov] (red) {Reduced-space optimisation \\ 
compute $\delta \boldsymbol{w}^{[j]}$ as: \\
Gauss--Newton: 
$\delta \boldsymbol{w}^{[j]}
= - \mathcal{H}_{\mathrm{GN}}^{-1}
 \dfrac{\partial J}{\partial \boldsymbol{w}} $ \\
L-BFGS:
$\delta \boldsymbol{w}^{[j]}
= - \alpha_j^{[j]} \mathcal{H}_{\mathrm{BFGS}}^{-1}
 \dfrac{\partial J}{\partial \boldsymbol{w}} $};

\node[box, left=of red] (anal) {Solution update \\ $\boldsymbol{w}^{[j+1]} = \boldsymbol{w}^{[j]} + \delta \boldsymbol{w}^{[j]}$  \\ $\boldsymbol{f}=\boldsymbol{f}^{(e)}+\mathbf{E}\boldsymbol{w}^{[j+1]}$};
\node[smallbox, left=of anal] (iter) {Check stopping criteria \\ Iterate $j+1$};

\draw[arr] (ens) -- (prop) node[lab]{};
\draw[arr] (prop) -- (cov) node[lab]{};
\draw[arr] (cov) -- (red) node[lab]{};
\draw[arr] (red) -- (anal) node[lab]{};
\draw[arr] (anal) -- (iter) node[lab]{};
\draw[arr] (iter) -- (prop) node[pos=0.15, lab]{};


\node[font=\footnotesize, above=2mm of ens] {Perturbations and initialization};
\node[font=\footnotesize, above=2mm of red, xshift=13mm] {solve in $\mathbb{R}^{N_{ens}}$};

\end{tikzpicture}
\caption{Schematic of the EnVar approach: ensemble generation, forward propagation, optimisation, solution update and iteration.}
\label{fig:envar_loop}
\end{figure}
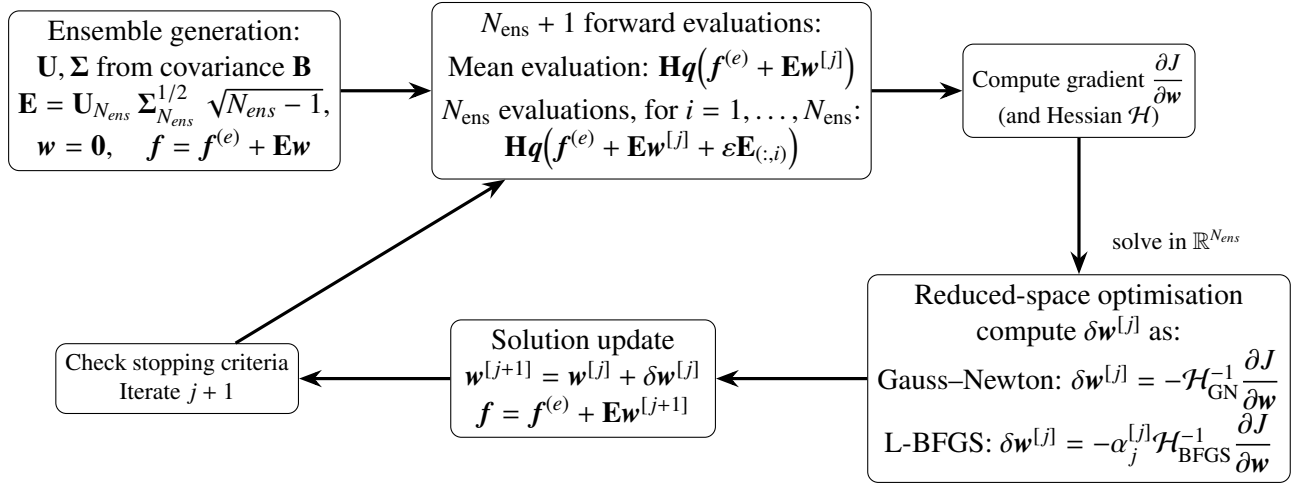

Figure \ref{fig:envar_loop} illustrates the main steps of the present iterative EnVar approach.

\subsubsection{Positioning with respect to other ensemble-based approaches}\label{sec:positioning}

This subsection situates the proposed EnVar scheme within the broader landscape of ensemble-based methods for inverse and optimisation problems, as developed in particular by the data assimilation community.

It may first be mentioned that EnVar methods are closely related to ensemble Kalman methods \cite{Evensen2009}, in particular as presented in \cite{Iglesias2013_ip} in the general context of inverse problems. In ensemble Kalman methods, the resolution of the inverse problem/minimisation of the considered cost function is carried out considering the original control vector (which would here amount to directly minimising $J(\boldsymbol{f})$ in \eqref{eq:Envarcost}), relying on ensemble-based estimations of covariance matrices. The EnVar approach contrasts with this strategy in that the minimisation is performed within the subspace spanned by the ensemble, as detailed above. This renders the EnVar approach more robust than ensemble Kalman methods with respect to the ensemble size $N_{ens}$ of the ensemble, as supported by \cite{Mons2016}. The EnVar approach may even rely on a single ensemble member (up to removing the factor $1/(N_{ens}-1)$ in \eqref{eq:ensemble_covariance}), putting aside the question of the representativeness of the optimal solution by such a small subspace, while ensemble Kalman methods are known to diverge for too small ensembles.

It is worth noting that the ensemble matrix $\mathbf{E}$ is held fixed throughout the minimisation procedure. This enables a consistent evaluation of the posterior statistics and avoids the ensemble-collapse phenomenon commonly observed in iterative ensemble methods where the ensemble is updated at each iteration, and where inflation procedures are typically required as a remedy \cite{Anderson1999_mwr}. An alternative to inflation is the consideration of iterative algorithms known as multiple data-assimilation schemes, which are discussed in, e.g., \cite{emerick2013,Chada2021}, but such approaches require the number of minimisation iterations to be specified in advance. A potential drawback of the present approach is that it places greater emphasis on the design of the ensemble matrix $\mathbf{E}$. However, in iterative ensemble methods where the ensemble is updated along the minimisation, the updated ensemble generally spans the same subspace as the initial one anyway \cite{Iglesias2013_ip}. In addition, as discussed above, it may be emphasized that the present way of generating the ensemble matrix $\mathbf{E}$ should be optimal to account for prior information for a given ensemble size. Finally, it is worth mentioning that, when considering $\varepsilon \ll 1$ in \eqref{eq:fdapprox}, the present EnVar scheme should better deal with nonlinearities in the model operator (i.e. the forced Navier-Stokes equations) compared to other ensemble methods that implicitly rely on $\varepsilon =1$ and the convexification of the objective function.

\section{Specification of the control objective and implementation details}\label{sec:control_objective_numerics}

In this section, the control objective for the present open-cavity configuration is first fully specified. Details about the EnVar and adjoint approaches are then provided. Finally, the numerical methods used to solve the Navier-Stokes equations and their adjoint counterpart are specified.

\subsection{Control objective and actuation strategy}\label{controlstrategy}

The aim of the flow-control procedure is here to reduce the intensity of the flow fluctuations, specifically targeting the shear layer spanning the cavity. More specifically, we want to minimise the velocity fluctuations at the four pointwise locations which are denoted by the yellow dots in Figure \ref{fig:cavity_tikz} and whose coordinates are given by \( x = [1/5, 2/5, 3/5, 4/5] \) and \(y=[0,0,0,0]\). Accordingly, the application of the observation operator $\mathbf{H}$ as introduced in \eqref{eq:Envarcost} to the flow state amounts to compute the fluctuating velocity $\boldsymbol{u}'=\boldsymbol{u}-\bar{\boldsymbol{u}}$, where $\bar{\boldsymbol{u}}$ refers to the time-averaged velocity, and to evaluate it at the specified pointwise locations, and also at chosen time instants. The target $\boldsymbol{y}$ is chosen as $\boldsymbol{0}$ (zero velocity fluctuation). Setting the covariance matrix $\mathbf{R}$ as the identity one, the first contribution in the cost function $J$ in \eqref{eq:Envarcost} may be written according to
\begin{equation}
    \left\| \boldsymbol{y} - \mathbf{H}\boldsymbol{q}(\boldsymbol{f})\right\|_{\mathbf{R}^{-1}}^2= \sum_{i,j}  \| \boldsymbol{u}'(\boldsymbol{x}_i,t_j,\boldsymbol{f}) \|^2,   
\end{equation}
where $\boldsymbol{x}_i$ refers to the location of the $i$th velocity sensor. The time instants $t_j$ are chosen to span a time period of $20$ convective times ($T=20$) after excluding an initial transient phase (also estimated around $20$ convective times). This ensures that the averaging window covers a sufficiently long time horizon that adequately captures the dynamics of the controlled flow.

The spatially-dependent forcing $\boldsymbol{f}_{\boldsymbol{u}}$ is chosen to act on the $y$-momentum equation only, namely $\boldsymbol{f}_{\boldsymbol{u}}(\boldsymbol{x})=(0,f(\boldsymbol{x}))^{\mathrm{T}}$. As discussed in Section \textsection~\ref{flow_conf}, the application of this forcing is restricted to the vicinity of the leading edge of the cavity, which, based on previous studies \cite{Barbagallo2009, Leclercq2019, Sipp2012}, should correspond to a subregion that allows efficient control of the flow downstream. This actuation subregion is illustrated in Figure \ref{fig:ellipse} and corresponds to the flow domain inside $[-0.32,0.15]\times[-0.60,0.15]$.

The first-guess for the (discrete) actuation forcing $\boldsymbol{f}^{(e)}$, used to initialize the optimisation procedures, is chosen as $\boldsymbol{f}^{(e)}=\boldsymbol{0}$, i.e. corresponding to an uncontrolled flow. The full cost function $J$ to minimise may thus be rewritten as
\begin{equation}\label{eq:specific_cost_function}
    J(\boldsymbol{f})=\frac{1}{2}\sum_{i,j}  \| \boldsymbol{u}'(\boldsymbol{x}_i,t_j,\boldsymbol{f}) \|^2+  \frac{1}{2} \left\| \boldsymbol{f}  \right\|_{\mathbf{B}^{-1}}^2,
\end{equation}
where the prior covariance matrix $\mathbf{B}$ is fully specified in the following subsection.

\subsection{Specification of the prior covariance matrix and EnVar parameters}\label{sec:prior_parameters}

\begin{figure}
    \centering
    \includegraphics[scale = 0.9]{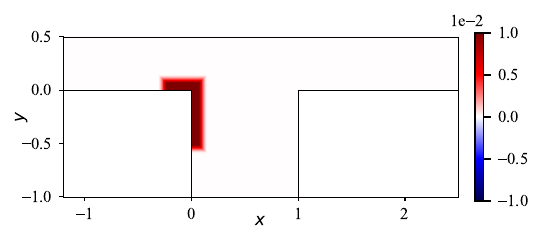}
    \caption{Control region and associated spatial distribution of the prior variance.}
    \label{fig:ellipse}
\end{figure}

The parameters of the prior covariance matrix \(\mathbf{B}\) in \eqref{eq:covarianceB} are chosen as follows. The streamwise and transverse correlation lengths are taken equal, \(a_x=a_y=0.07\), in order not to introduce a preferential alignment of the ensemble perturbations. This value is much smaller than the characteristic length of the problem, \(D\), by more than one order of magnitude, while remaining sufficiently large to avoid excessively localized actuation patterns. The cross-correlation parameter is set to \(c_{xy}=\infty\), so that no preferential oblique correlation is imposed. Concerning the variance parameter $\sigma_i^2$, whose inverse locally weights the penalization of the intensity of the actuation $\boldsymbol{f}$ (see also \eqref{eq:specific_cost_function}), it is chosen to vary in space as follows. As illustrated by Figure \ref{fig:ellipse}, the local variance reaches a maximum value of \(\sigma_{\max}^{2}=0.01\) in the central part of the actuation region, corresponding to a significant level of penalization, and decreases smoothly towards its boundaries. Outside of this region, $\sigma_i^2=0$, corresponding to an infinite penalization of the actuation. It may be emphasized that the aim of the present study is to demonstrate the potentialities of the EnVar approach in tackling high-dimensional open-loop flow-control problems and to perform comparisons with the adjoint technique, and a detailed investigation about the sensitivity of the optimisation results with respect to the above parameters appears here out of scope. Some supplementary discussion on the choice of the correlation-length parameters is still provided in \ref{appa}.

\begin{figure}
    \centering
    \begin{subfigure}{0.45\textwidth}
        \centering
        \includegraphics[width=.9\textwidth]{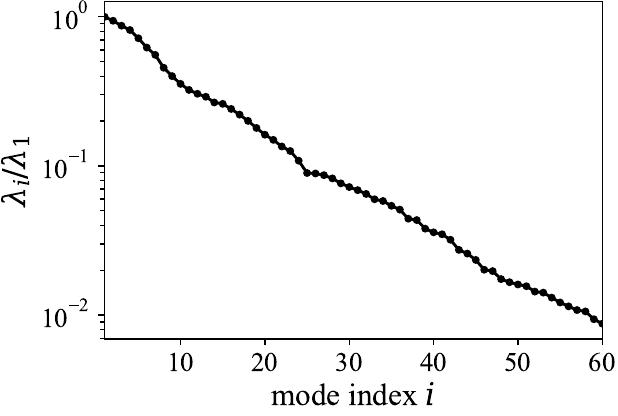}
        \caption{Eigenvalue spectrum.}
        \label{fig:B_eigenvalue_spectrum}
    \end{subfigure}
    \begin{subfigure}{1\textwidth}
        \centering
        \includegraphics[width=\textwidth]{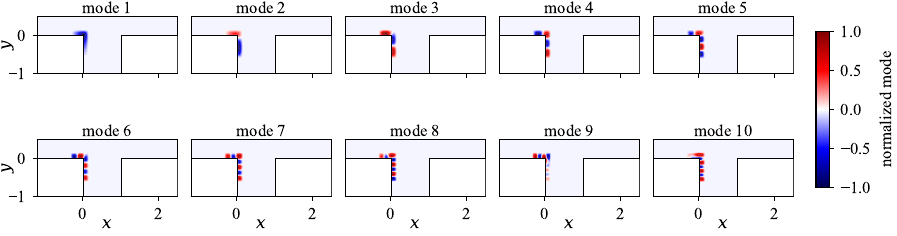}
        \caption{First ten eigenvectors.}
        \label{fig:B_eigenvectors_first10}
    \end{subfigure}
    \caption{Spectrum and leading eigenvectors of the covariance matrix \(\mathbf{B}\). The eigenvectors define the dominant directions retained in the ensemble subspace.}
    \label{fig:B_spectrum_eigenvectors}
\end{figure}

The eigenvalue spectrum of the covariance matrix \(\mathbf{B}\), together with the first ten eigenvectors, is shown in Figure~\ref{fig:B_spectrum_eigenvectors}. The rapid decay of the eigenvalues indicates that the dominant spatial structures of the covariance can be represented by a reduced number of modes, while the associated eigenvectors provide smooth and spatially coherent forcing directions, corresponding to an almost constant field for the dominant mode to more and more oscillatory spatial variations for the following ones. This rapid decay of the eigenvalue spectrum justifies considering only a few of these eigenvectors in the EnVar approach (see Section~\ref{sec:ensemblematrix}), while it may be noted that the actuation region specified in Section \ref{controlstrategy} corresponds to more than $3\cdot 10^3$ degrees of freedom (thus corresponding to the effective size of the discrete forcing $\boldsymbol{f}$) in the following test cases. The influence of the number $N_{ens}$ of kept eigenvectors will be investigated in the following.

The parameter $\varepsilon$ to perform the finite-differences in \eqref{eq:fdapprox} is chosen as $10^{-4}$. This value was chosen as a compromise between the accuracy in the evaluation of the matrix $\mathbf{Y}$ and the avoidance of cancellation by round-off errors. The evaluation of $\mathbf{Y}$ is not expected to be significantly impacted by reasonable changes around this value, while keeping in mind that a majority of ensemble techniques implicitly rely on $\varepsilon=1$, amounting to convexify the cost function $J$ around the current estimate of $\boldsymbol{f}$, as discussed in Section \ref{sec:positioning}.

\subsection{Gradient regularization in the adjoint approach}

As the present cavity configuration exhibits sharp corners, in particular at the leading edge, which corresponds to a highly sensitive region \cite{Sipp2012}, the gradient \eqref{eq:gradient_forcing} in the adjoint approach may happen to be highly concentrated in this same region, possibly leading to non-smooth updates in the actuation forcing during the minimisation process. To counteract this, we here apply a $H^1$-like regularization of this gradient as discussed in \cite{Protas2004_jcp}. Namely, from the gradient $\partial \mathscr{L}/\partial f$, here considering the continuous counterpart of \eqref{eq:gradient_forcing} for the sake of simplicity, we infer a more regular gradient $(\partial \mathscr{L}/\partial f)|^{H^1}$ by solving 
\begin{equation}
\left(
\mathbb{I}
-
l_{\mathrm{sob}}^2 \Delta
\right)
\frac{\partial \mathscr{L}}{\partial f}\Big|^{H^1}
=
(1+l_{\mathrm{sob}}^2)
\frac{\partial \mathscr{L}}{\partial f},
\label{lsob}
\end{equation}
where the parameter $l_{\mathrm{sob}}$ corresponds to a filter length. Its value is chosen to be comparable to the correlation length scales as specified for the prior covariance matrix (see Section \ref{sec:prior_parameters}). Here, the filter length is set to $l_{\mathrm{sob}} = 0.02$, which is comparable to the values of the correlation-length parameters for the prior covariance matrix $\mathbf{B}$ (see Section \ref{sec:prior_parameters}). While a range of values is admissible, excessively large filter lengths would overly smooth the gradient field, whereas excessively small values would be insufficient to regularize localized singularities.

\subsection{Numerical methods and boundary conditions}\label{sec:BC_numerical_methods}

The incompressible forced Navier-Stokes equations \eqref{eq:discrete_NS} and their adjoint counterpart \eqref{eq:adjoint_discrete} are solved using the finite element method. All equations considered in this study are first reformulated in variational form and then spatially discretised on a triangular mesh. The triangulation is generated using the open-source software FreeFEM \cite{Hecht2012}. The velocity and pressure fields are discretised with Taylor–Hood elements ($P_2,P_2,P_1$), employing quadratic $P_2$ basis functions for velocity components and linear $P_1$ functions for pressure.
All discrete matrices, obtained by projecting the variational formulations onto the finite element basis, are assembled in FreeFEM using a domain-decomposition parallelisation strategy and interfaced with the PETSc library for the solution of large sparse linear systems. The matrices are sparse, and their inverses are computed with the UMFPACK library, which provides a sparse direct LU solver.
A fractional-step method \cite{codina2001} is used to solve the incompressible Navier-Stokes equations. Time integration is performed
in an implicit way based on a second-order backward finite-difference scheme.
Boundary conditions for the Navier-Stokes equations are as follows (see also Figure~\ref{fig:cavity_tikz} and Table \ref{tabgeometry}). A uniform velocity ($u_x = 1$, $u_y = 0$) is imposed at the inlet. A symmetry boundary condition ($\frac{\partial u_x}{\partial y} = 0$, $u_y = 0$) is applied at the top boundary, as well as along ($-L_1 < x < -L_{1,slip}, y=0$) and ($D+L_{2,slip} < x < D+L_2, y=0$). Within the cavity region, for all $x \in [-L_{1,slip}, L_{2,slip}]$, a no-slip condition ($u_x = 0$, $u_y = 0$) is enforced. At the outlet boundary, ($Re^{-1}\frac{\partial u_x}{\partial x} -p = 0$, $\frac{\partial u_y}{\partial x} = 0$) is imposed.
The adjoint Navier-Stokes equations are derived and solved following the discrete adjoint approach. A checkpointing strategy is applied to avoid storing the forward solution at all time steps. The same numerical methods were used in \cite{Mons2022}.

\section{Results}\label{sect5}

In this section, the methodologies of Section \ref{sect3} are applied to the identification of an optimal actuation in the sense of Section \ref{controlstrategy}, first considering a moderate Reynolds number of $6250$. In particular, results obtained with the EnVar approach are compared with those obtained with the adjoint one. A more challenging regime corresponding to $Re = 14000$ is then examined to evaluate the robustness of the EnVar methodology.


\subsection{Moderate Reynolds number -- $Re=6250$}\label{Re6250}

\begin{figure}
    \centering
    \begin{subfigure}{0.46\textwidth}
        \centering
        \includegraphics[scale = 1]{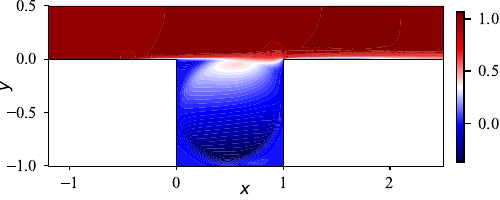}
        \caption{Mean flow.}
        \label{fig:base_meana}
    \end{subfigure}
    \hfill  
    \begin{subfigure}{0.46\textwidth}
        \centering
        \includegraphics[scale = 1]{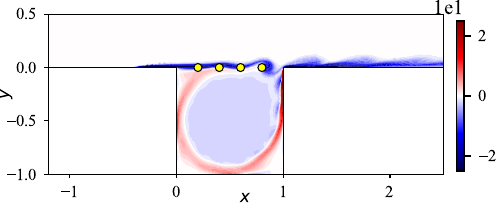}
        \caption {Vorticity snapshot.} 
        \label{fig:base_vorticity}
    \end{subfigure}
    \caption{Mean streamwise velocity field (a) and instantaneous vorticity field (b) for $Re=6250$.}
    \label{fig:base_mean}
\end{figure}

\begin{figure}
    \centering
    \begin{subfigure}{0.337\textwidth}
        \centering
        \includegraphics[width=\linewidth]{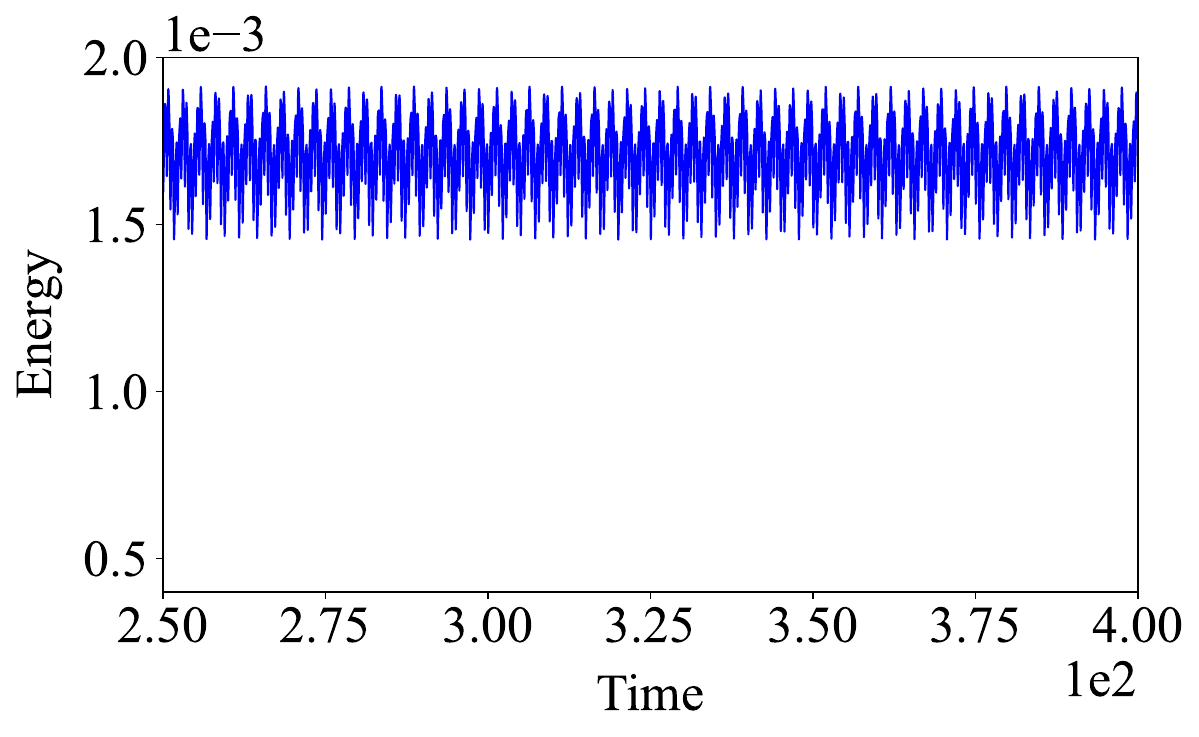}
        \caption{Time evolution of the global kinetic energy.}
        \label{fig:energyunconaa}
    \end{subfigure}
    \hfill
    \begin{subfigure}{0.337\textwidth}
        \centering
        \includegraphics[width=\linewidth]{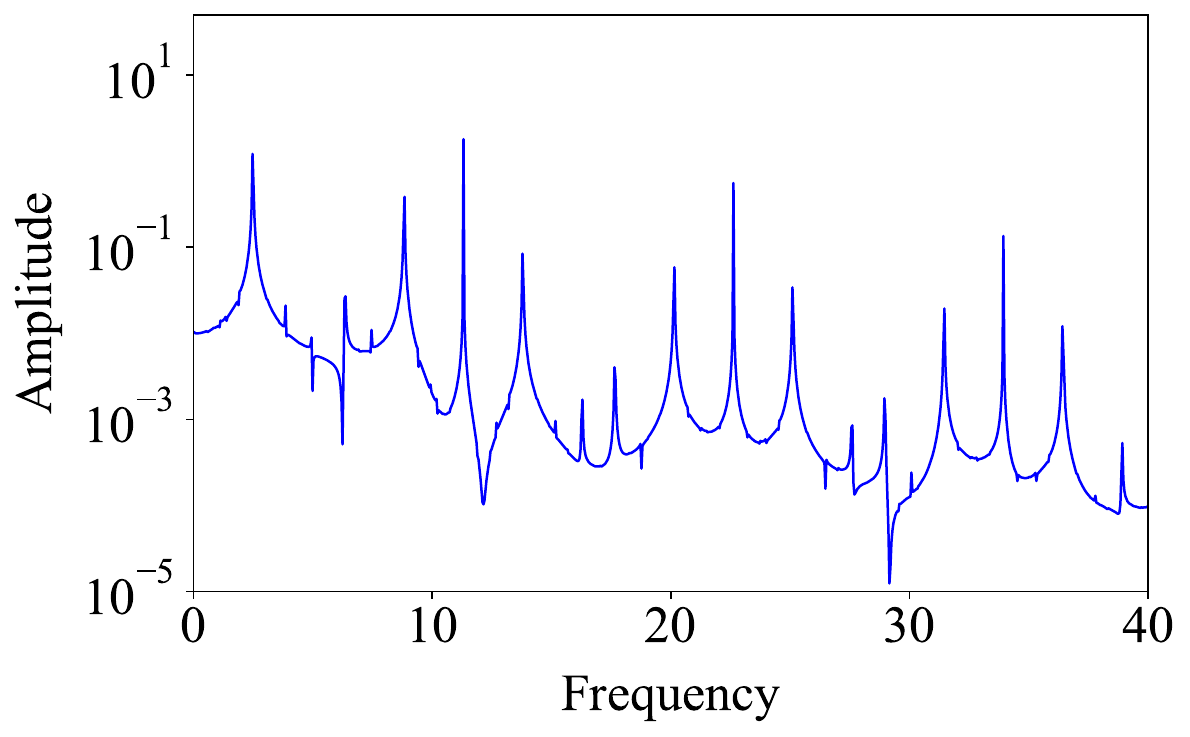}
        \caption{Frequency spectra.}
        \label{fig:energyuncona}
    \end{subfigure}
    \hfill
    \begin{subfigure}{0.305\textwidth}
        \centering
        \includegraphics[width=\linewidth]{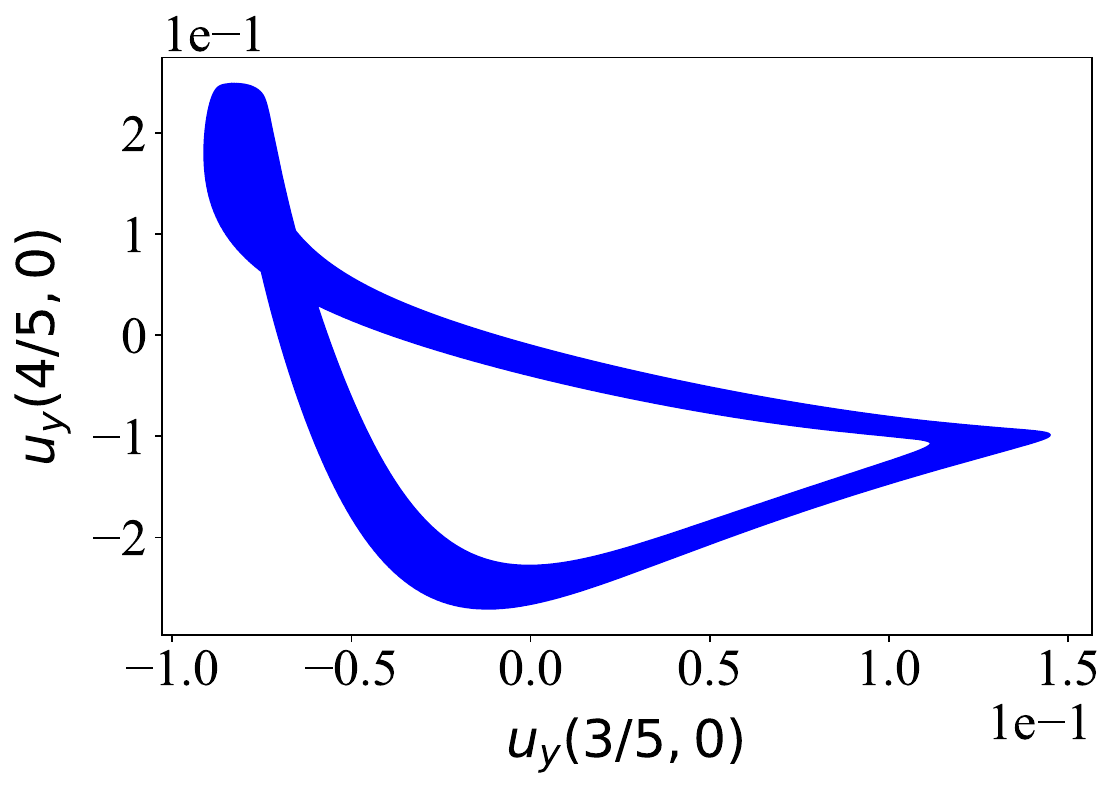}
        \caption{Trajectory in phase space.}
        \label{fig:energyunconb}
    \end{subfigure}
    \caption{Characterisation of the uncontrolled quasi-periodic flow dynamics at $Re=6250$. The time evolution of the global kinetic energy (a), the corresponding frequency spectra (b), and the trajectory constructed from $y$-velocity located in the shear layer at 
$y = 0$ and at streamwise positions $x = \tfrac{3}{5}$, $x = \tfrac{4}{5}$ (c) are reported.}
    \label{fig:energyuncon2}
\end{figure}

The present open-cavity flow configuration is first examined for $Re=6250$. Figure \ref{fig:base_mean} illustrates the streamwise velocity component of the mean flow and a vorticity snapshot in the uncontrolled case. The mean flow (Figure \ref{fig:base_meana}) reveals smooth recirculation patterns within the cavity and the development of the shear layer across the opening of the cavity. In particular, it shows a thickened boundary layer near the downstream edge, reflecting the persistent influence of oscillatory motions averaged over time. The vorticity snapshot (Figure \ref{fig:base_vorticity}) allows us to identify regions of strong rotational motion and the development of Kelvin–Helmholtz instabilities. As in Figure \ref{fig:cavity_tikz}, the yellow dots in this figure refer to the locations at which velocity fluctuations are measured to evaluate the cost function $J$ in \eqref{eq:specific_cost_function}. Figure~\ref{fig:energyuncon2} characterizes the quasi-periodic dynamics of this uncontrolled flow. In particular, Figure~\ref{fig:energyunconaa} reports the time evolution of the global kinetic energy defined as
\begin{equation}
\begin{aligned}
E_k= \int^T_0 \int_{\Omega}(u_x^{'2}+u_y^{'2}) \,d\boldsymbol{x}dt,  
\end{aligned}
\label{kinenergy}
\end{equation}
while Figure~\ref{fig:energyuncona} shows the corresponding Fourier spectrum, allowing the dominant frequencies of the quasi-periodic dynamics to be identified. This quasi-periodic behaviour is further illustrated in Figure \ref{fig:energyunconb}, which reports a phase portrait based on the transverse velocity at two locations that correspond to the two most downstream yellow dots in Figure \ref{fig:base_vorticity}. This flow exhibits a dominant oscillation at a frequency of $\omega = 11.31$. A lower-frequency component at $\omega = 2.49$ also appears, and the remaining peaks in the spectrum in Figure~\ref{fig:energyuncona} arise from interactions between these two primary frequencies. Further insights about the interactions between these fundamental frequencies can be found in \cite{Leclercq2019}. 

\subsubsection{Identification of optimal actuation through EnVar and adjoint approaches}\label{sec:adjoint_EnVar}

We now proceed with the determination of an optimal actuation forcing in order to minimise the intensity of the flow fluctuations in the sense of the cost function $J$ in \eqref{eq:specific_cost_function}. The minimisation of this cost function is performed here by relying on either the adjoint or EnVar approach. In addition, regarding the EnVar approach, we consider the use of either the Gauss–Newton or L-BFGS descent method, which amounts to a total of three optimisation procedures. The EnVar procedures rely on an ensemble of size \(N_{\mathrm{ens}}=20\). This appears to be a reasonable choice given the rapid decay of the eigenvalue spectrum of Figure \ref{fig:B_eigenvalue_spectrum} discussed in Section \ref{sec:prior_parameters} ($\lambda_{20}/\lambda_1 \simeq 0.1$), and other values of \(N_{\mathrm{ens}}\) will be investigated in Section \ref{sec:sensitivity_nens}.

\begin{figure}
    \centering
    \begin{subfigure}{0.48\textwidth}
        \centering
        \includegraphics[width=\textwidth]{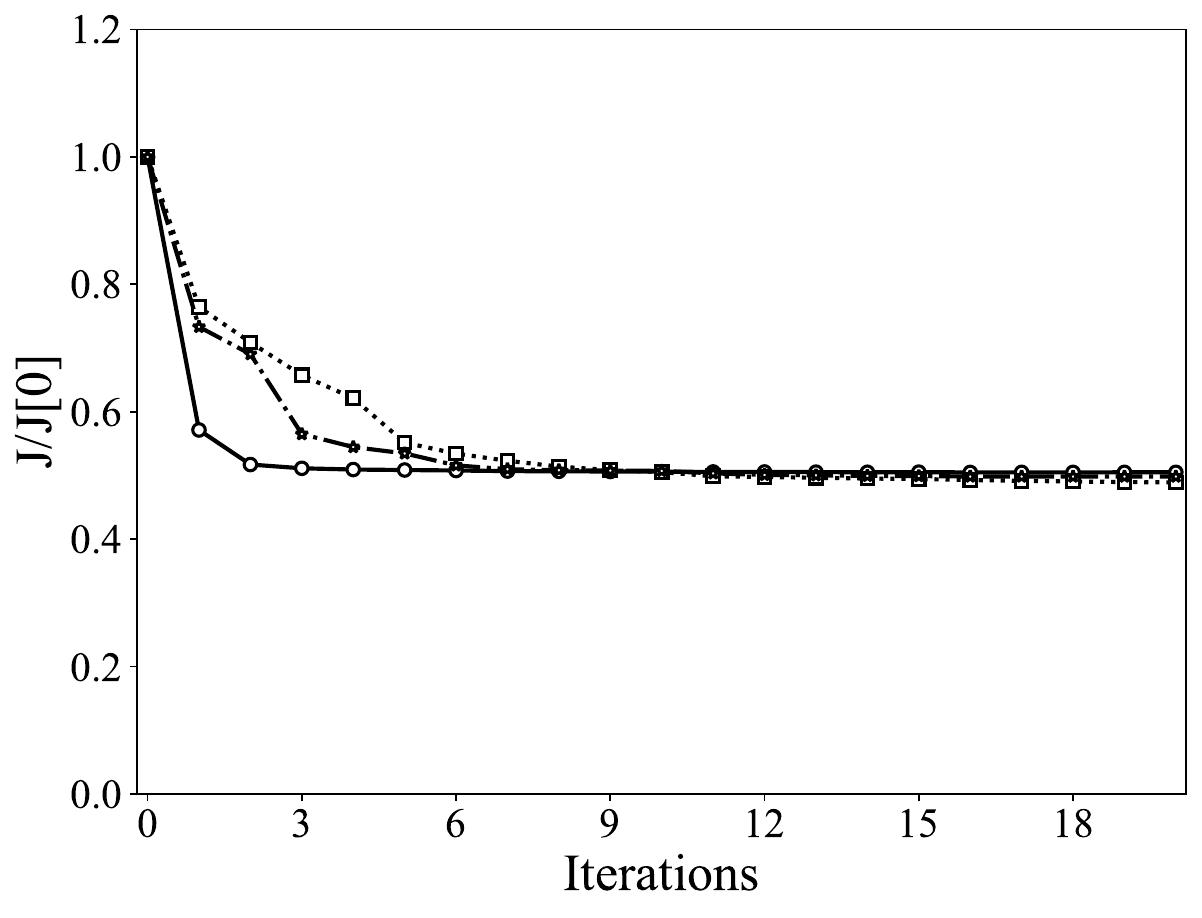}
        \caption{Total normalised cost function for the three optimisation strategies.}
        \label{fig:costfuncdecfreq0_total}
    \end{subfigure}
    \hfill
    \begin{subfigure}{0.48\textwidth}
        \centering
        \includegraphics[width=\textwidth]{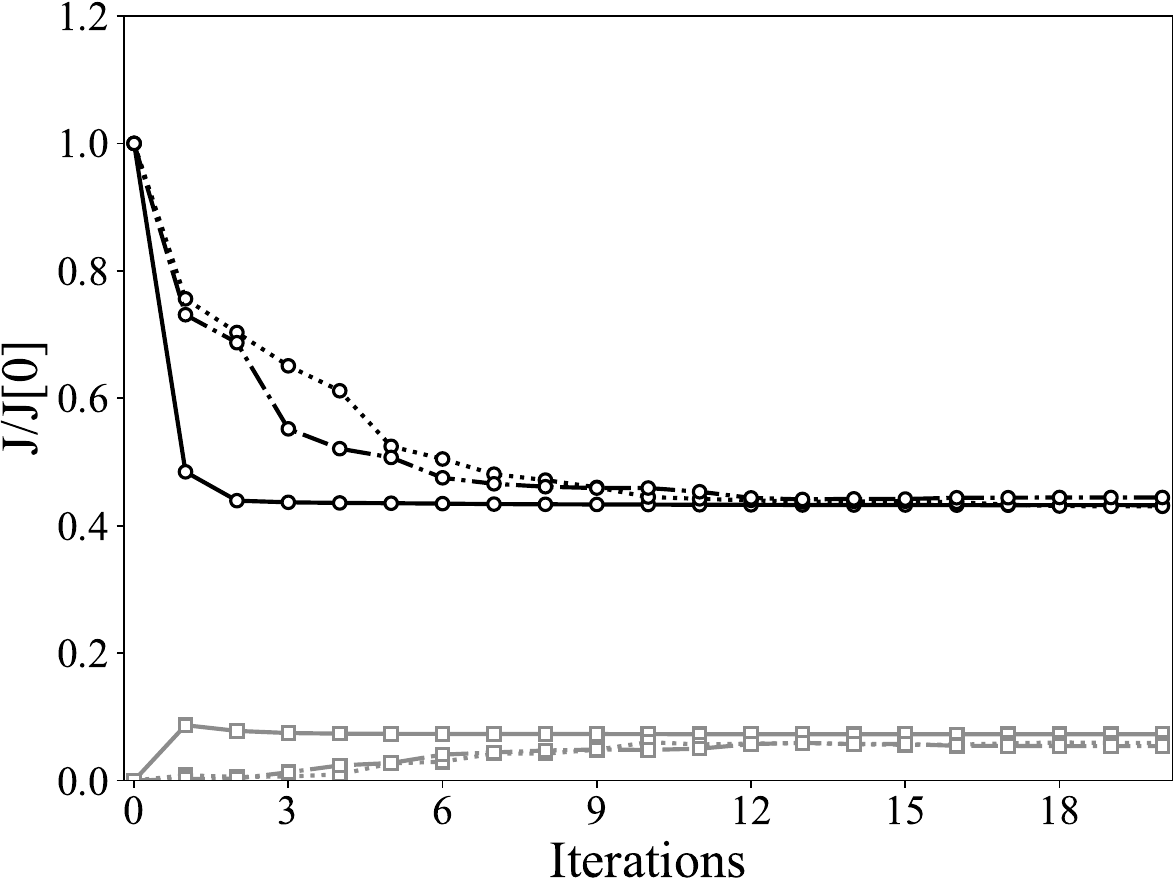}
        \caption{Observation term (black lines) and regularisation term (grey lines).}
        \label{fig:costfuncdecfreq0_components}
    \end{subfigure}

    \caption{Evolution (a) and decomposition (b) of the cost function for the EnVar Gauss–Newton ($-$), the EnVar L-BFGS ($-\cdot-$), and the adjoint-based optimisation algorithm ($\cdots$), at $Re=6250$.}
    \label{fig:costfuncdecfreq0}
\end{figure}

Figure \ref{fig:costfuncdecfreq0} shows the convergence of the cost function $J$ for the three optimisation cases: Figure \ref{fig:costfuncdecfreq0_total} reports the evolution of the total cost function, while Figure \ref{fig:costfuncdecfreq0_components} displays the separate contributions of the fluctuating velocity ($\| \boldsymbol{u}'(\boldsymbol{x}_i,t_j,\boldsymbol{f}) \|^2$) and regularization ($\left\| \boldsymbol{f}  \right\|_{\mathbf{B}^{-1}}^2$) terms. It appears that the three cases reach very similar final values of the cost function after a significant reduction within the first few iterations of the optimisation procedure. Interestingly, the convergence history in the EnVar-L-BFGS case is closer to that for the adjoint approach (which also relies on L-BFGS) compared to the EnVar-Gauss-Newton case. Figure \ref{fig:costfuncdecfreq0_components} indicates that the intensity of velocity fluctuations (as quantified in the cost function $J$ and measured at the considered locations) has been decreased to approximately 40\% of its initial value (corresponding to the uncontrolled flow), which confirms the efficacy of the optimal actuation in all cases. It may be noted that this result is obtained for a specific choice for the prior variance $\sigma^2_i$, which determines the compromise between fluctuation attenuation and control regularisation. While the identification of an optimal compromise between attenuation and regularisation, i.e. of an optimal $\sigma^2_i$, appears outside the scope of the present study, as discussed in Section \ref{sec:prior_parameters}, this could constitute an avenue for future work.

\begin{figure}
    \centering
    \includegraphics[scale = 0.9]{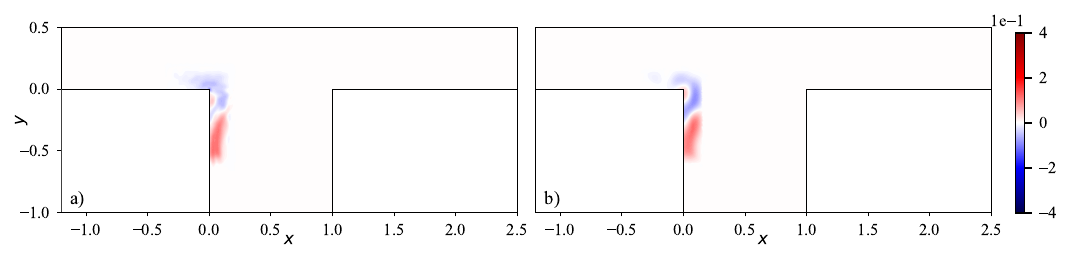}
        \centering
    \includegraphics[scale = 0.9]{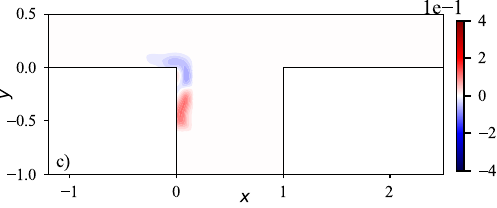}
    \caption{Optimal forcing obtained with the adjoint method (a), EnVar Gauss--Newton (b), EnVar L-BFGS (c), at $Re=6250$.}
    \label{fig:ellipse12}
\end{figure}

Figure \ref{fig:ellipse12} reports the optimal actuation forcing obtained in the three cases. These forcing distributions are highly similar, while above discussions confirm that they induce the same performances in terms of reduction of the velocity fluctuations. This demonstrates that the EnVar approach can satisfactorily infer a spatially dependent quantity while yielding results consistent with those obtained using the adjoint approach. This finding is here valid regardless of whether the EnVar approach relies on the Gauss-Newton or L-BFGS descent method, which will be further investigated in the following.

\subsubsection{Analysis of the controlled flow}

\begin{figure}
    \centering
    \begin{subfigure}{0.337\textwidth}
        \centering
        \includegraphics[width=\linewidth]{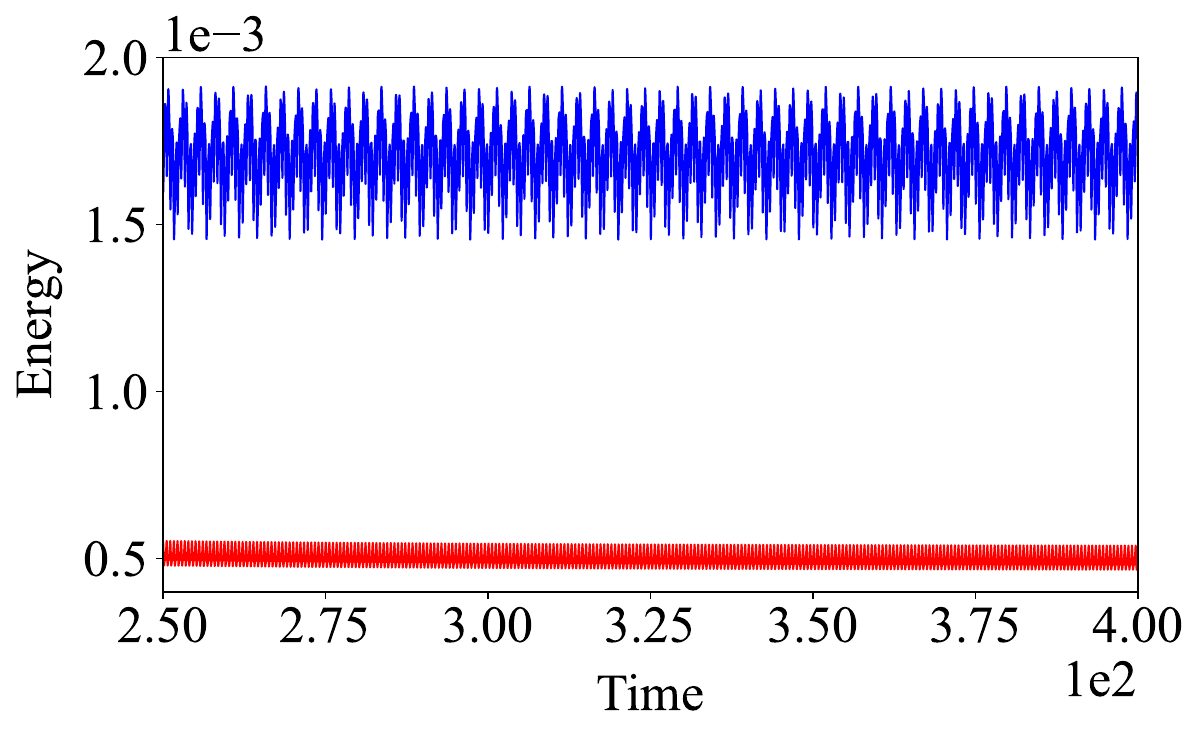}
        \caption{Time evolution of the global kinetic energy.}
        \label{fig:energyunconava}
    \end{subfigure}
    \hfill
    \begin{subfigure}{0.337\textwidth}
        \centering
        \includegraphics[width=\linewidth]{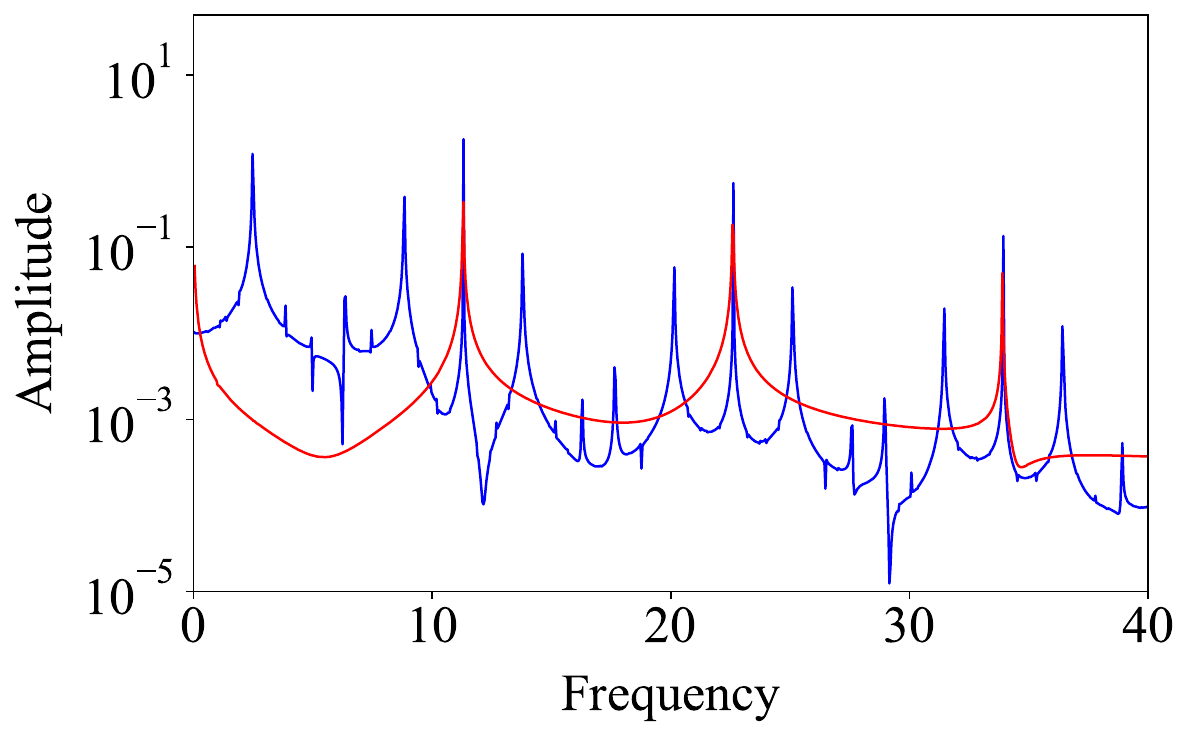}
        \caption{Frequency spectra.}
        \label{fig:energyunconav}
    \end{subfigure}
    \hfill
    \begin{subfigure}{0.305\textwidth}
        \centering
        \includegraphics[width=\linewidth]{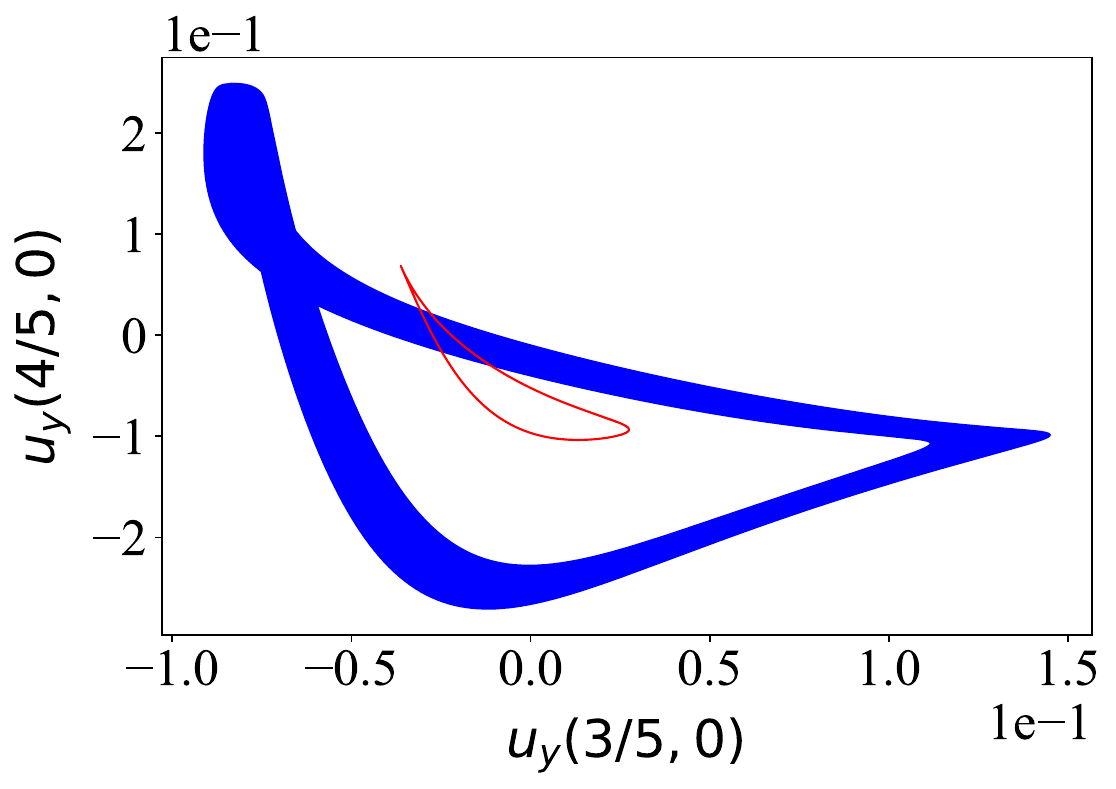}
        \caption{Trajectory in phase space.}
        \label{fig:energyunconbv}
    \end{subfigure}
    \caption{Characterisation of quasi-periodic flow dynamics under controlled and uncontrolled conditions, at $Re=6250$. The uncontrolled flow is shown in blue, whereas the controlled flow obtained with the Gauss--Newton EnVar method is shown in red. The time evolution of the global kinetic energy (a), the corresponding frequency spectra (b), and the trajectory constructed from $y$-velocity located in the shear layer at 
$y = 0$ and at streamwise positions $x = \tfrac{3}{5}$, $x = \tfrac{4}{5}$ (c) are reported.}
    \label{fig:energyuncon2vv}
\end{figure}

In this subsection, the dynamics of the controlled flow are further investigated. As the optimisation procedures discussed in Section \ref{sec:adjoint_EnVar} produced very similar solutions, we here only report results obtained through the EnVar-Gauss-Newton approach, with almost identical findings for the two other optimisation cases. Figure \ref{fig:energyuncon2vv} compares the controlled flow (red lines) with its uncontrolled counterpart (blue lines). In particular, Figure \ref{fig:energyunconava} presents the time evolution of the global kinetic energy, computed as in \eqref{kinenergy}, revealing a reduction of more than threefold in the controlled case. This confirms the global effect of the actuation forcing despite the consideration of only a few locations to evaluate the cost function $J$. The corresponding frequency spectrum is presented in Figure \ref{fig:energyunconav}. As previously discussed, in the uncontrolled case, the spectrum exhibits multiple peaks that correspond to the quasi-periodic nature of the flow and reflect the presence of several interacting oscillatory modes. Upon application of the optimised steady actuation forcing, the flow is dominated by a single fundamental frequency and its harmonics, signaling a transition to predominantly periodic behaviour and a simplification of the overall flow dynamics. Figure \ref{fig:energyunconbv} further highlights this transition by showing the trajectory in phase space. Here, the controlled flow exhibits a tightly constrained, periodic orbit with substantially reduced fluctuation amplitudes compared to the uncontrolled case.

\begin{figure}
    \centering
    \includegraphics[scale = 0.9]{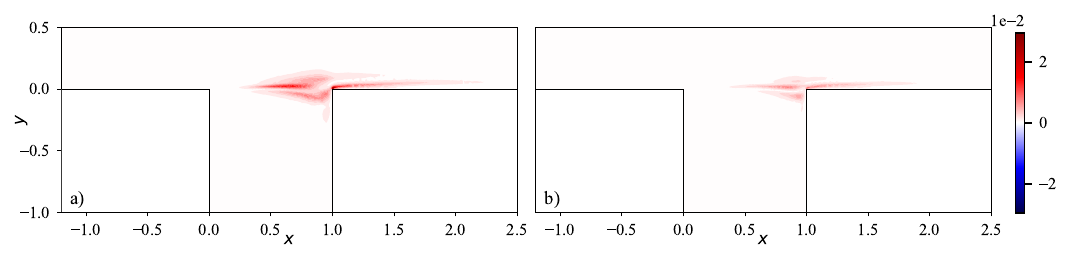}
    \caption{Reynolds stress component $\langle u'u' \rangle$ at $Re=6250$. Uncontrolled flow (a) and controlled flow obtained with the Gauss--Newton EnVar method (b).}
    \label{fig:STATS_uu}
\end{figure}

\begin{figure}
    \centering
    \includegraphics[scale = 0.9]{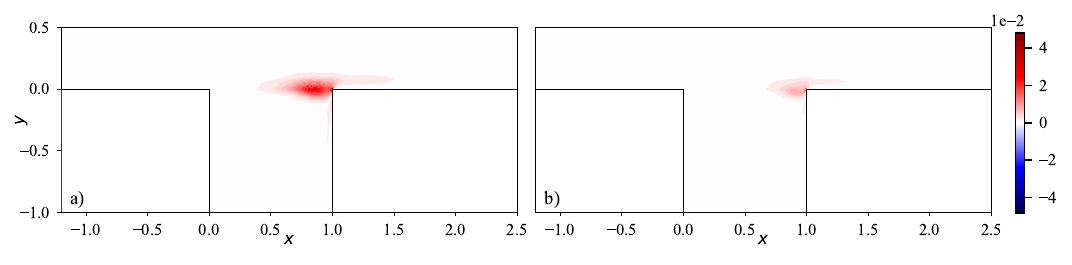}
    \caption{Reynolds stress component $\langle v'v' \rangle$ at $Re=6250$. Uncontrolled flow (a) and controlled flow obtained with the Gauss--Newton EnVar method (b).}
    \label{fig:STATS_vv}
\end{figure}

The spatial distribution of statistical quantities such as the Reynolds stresses components, shown in Figures \ref{fig:STATS_uu} and \ref{fig:STATS_vv}, provides additional insight into the effect of the control. The streamwise $\langle u'u' \rangle$ and transverse $\langle v'v' \rangle$ components both display a clear reduction in magnitude across the whole shear layer spanning the opening of the cavity. This is particularly noteworthy for the transverse and dominant component $\langle v'v' \rangle$ (Figure \ref{fig:STATS_vv}). Together, these results further demonstrate that the identified actuation forcing effectively stabilizes the flow, converting a quasi-periodic system into a periodic one, while significantly attenuating velocity fluctuations throughout the cavity.


\subsubsection{Sensitivity with respect to the number of ensemble members}\label{sec:sensitivity_nens}

\begin{figure}
    \centering
    \includegraphics[scale = 0.4]{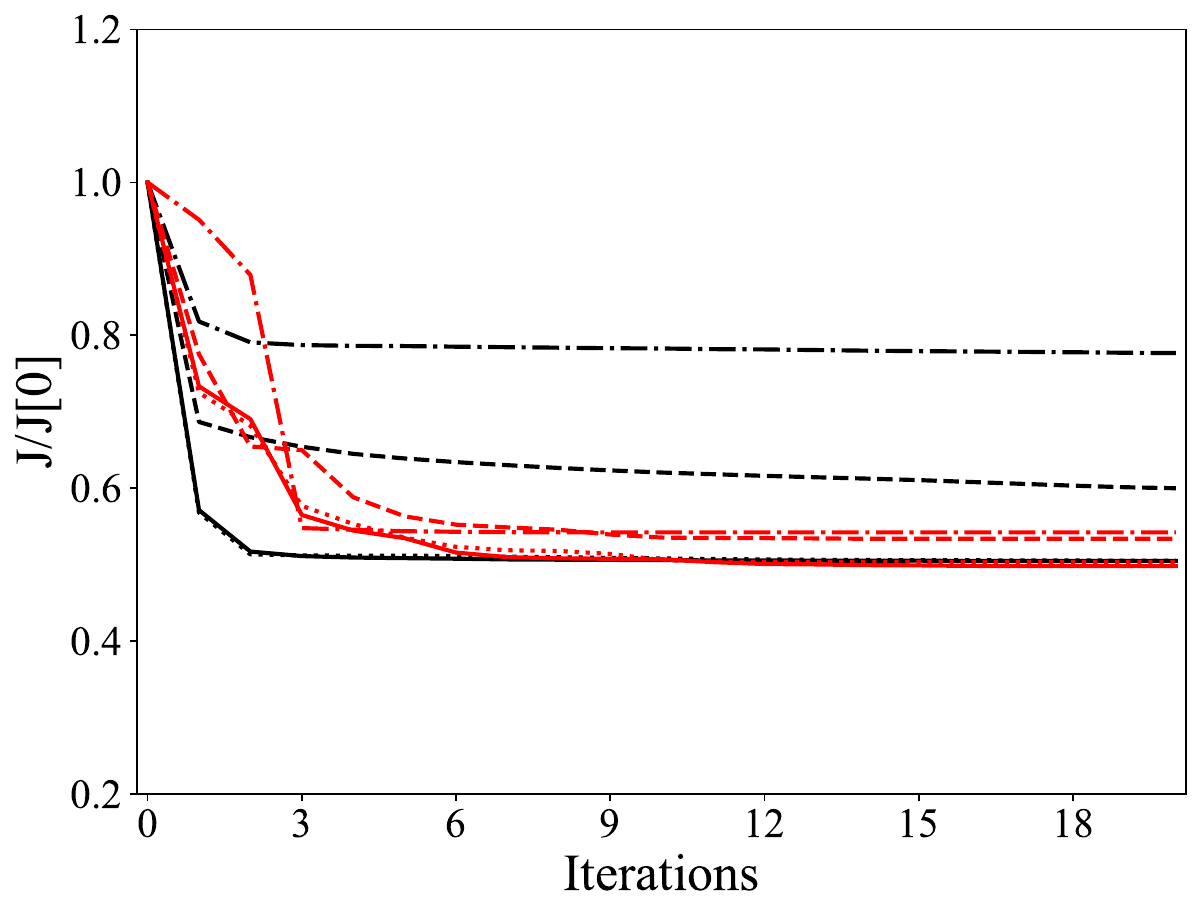}
    \caption{Evolution of the cost function for the EnVar-based optimisation algorithm for different numbers of ensemble members at $Re=6250$. Four ensemble sizes are considered (5, 10, 20, and 40 members), represented by $-\cdot-$, $--$, ---, and $\cdots$, respectively. Results are reported for both the EnVar Gauss–Newton method (black lines) and the EnVar L-BFGS method (red lines). }
    \label{fig:costfuncdecfreq01}
\end{figure}

\begin{figure}
    \centering
    \includegraphics[scale = 0.9]{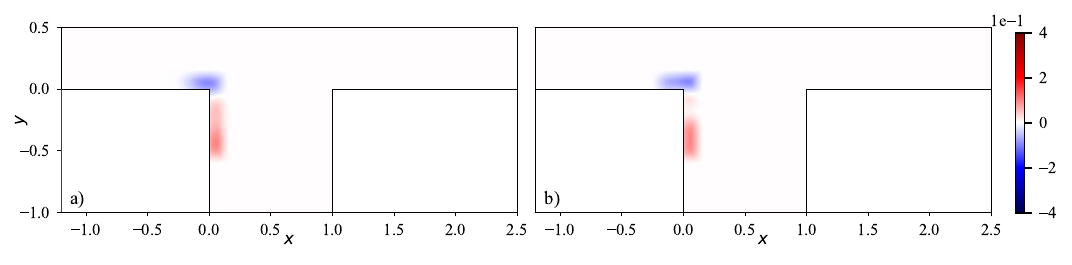}
        \includegraphics[scale = 0.9]{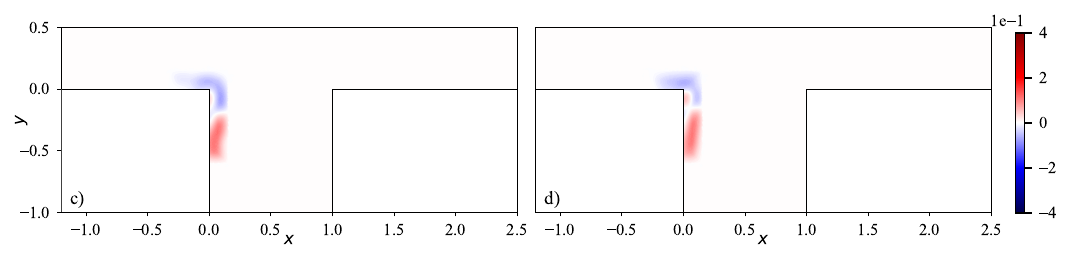}
    \caption{Optimal forcing obtained at  $Re=6250$ with EnVar L-BFGS method using 5 ensemble members (a), 10 ensemble members (b), 20 ensemble members (c),  40 ensemble members (d).}
    \label{fig:ellipse10}
\end{figure}

The size of the ensemble $N_{ens}$ is an important parameter in the EnVar approach, determining the dimension of the subspace in which the optimisation is performed and directly impacting the computational cost of the procedure (see Sections \ref{sec:EnVar_scheme} and \ref{sec:computational_cost}). Consequently, a sensitivity analysis with respect to $N_{ens}$ is performed, whose results are reported in Figures \ref{fig:costfuncdecfreq01} and \ref{fig:ellipse10}. Figure \ref{fig:costfuncdecfreq01} shows the evolution of the cost functional $J$ over the optimisation iterations for $N_{ens}=5,10,20,40$, considering the use of both the Gauss--Newton and L-BFGS descent methods. For all ensemble sizes, both approaches yield a rapid initial decrease of the cost function, with a substantial reduction already after the first update, confirming the effectiveness of the ensemble-based approximation of the gradient. For larger ensembles, namely $N_{ens}=20$ (as already discussed above) and $N_{ens}=40$, the two optimisation strategies exhibit very similar convergence behaviour: although the L-BFGS method converges more gradually, it ultimately reaches the same optimal value as the Gauss--Newton approach, with both methods showing minimal oscillations and nearly identical final costs. This indicates that, when the ensemble dimension is sufficiently large, both methods access similar descent directions within the reduced subspace. In contrast, clear differences emerge for smaller ensemble sizes. For $N_{ens}=10$ and especially for $N_{ens}=5$, the Gauss--Newton method displays a slower convergence, with larger discrepancies in the cost value. In these cases, the L-BFGS algorithm performs markedly better, achieving a lower cost and a more robust minimisation of the objective function. This behaviour suggests that, for small ensembles, the reduced space optimisation is better handled by the L-BFGS update and its line-search strategy than by the direct Gauss--Newton step.

Figure \ref{fig:ellipse10} further examines the spatial structure of the optimal forcing obtained for the different ensemble sizes within the L-BFGS method. The fields associated with $N_{ens}=20$ and $N_{ens}=40$ are almost indistinguishable, displaying a highly coherent and smooth actuation pattern. This indicates that $N_{ens}=20$ already corresponds to converged results with respect to the ensemble size. The forcing obtained with $N_{ens}=10$ remains very similar to the large-ensemble solutions, capturing most of the dominant spatial features, but with some discrepancies near the cavity edge. The forcing associated with $N_{ens}=5$ exhibits an even more noticeable loss of spatial richness, the ensemble being formed by only the $5$ most dominant modes in Figure \ref{fig:B_spectrum_eigenvectors}. This solution still appears in line with the largest-scale features of the forcings obtained with larger ensembles. Furthermore, it may be observed that the corresponding value of the cost function is not significantly higher than for the other cases (red curves in Figure \ref{fig:costfuncdecfreq01}). A further analysis of these different forcings is provided in \ref{appb}.

Overall, these results highlight a key strength of the proposed EnVar approach: even with very small ensemble sizes, a coherent and physically meaningful forcing can be extracted, and flow optimisation is achieved within very few evaluations of the objective function, given that a robust descent method such as L-BFGS is employed. While larger ensembles allow finer flow structures to be captured, the global forcing structure is already recovered with a limited number of ensemble directions — a property not shared by ensemble Kalman filters \cite{katzfuss2016understanding}, which generally require substantially larger ensembles to produce reliable updates. 

\subsection{Higher Reynolds number regime --   $Re=14000$}\label{Re14000}


Having assessed the EnVar approach in the quasi-periodic regime, demonstrating its consistency with the adjoint approach and its robustness with respect to the size of the ensemble in this case, attention is now turned to a more challenging configuration characterised by a positive largest Lyapunov exponent, indicating chaotic flow. In such regimes, small perturbations grow exponentially in time, leading to strong sensitivity to initial conditions, which may compromise gradient accuracy and numerical conditioning over long time horizons. This is especially the case when relying on linearized models as in the adjoint approach, although some remedies have been proposed in previous studies \cite{Wang2013_jcp}.
The configuration chosen is therefore the open-cavity flow at a higher Reynolds number of $Re=14000$. The objective of this test case is to assess the robustness of the EnVar approach in this more demanding setting, considering the same time horizon as in the case $Re=6250$, which is here comparable to one Lyapunov time. We also aim to verify that the ensemble size required for optimisation remains within acceptable bounds, and to examine the effects of optimised forcing obtained through the EnVar framework on the flow dynamics.

\begin{figure}
    \centering
    \begin{subfigure}{0.46\textwidth}
        \centering
        \includegraphics[scale = 1]{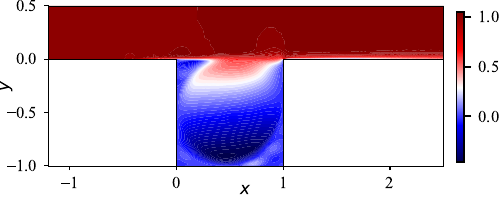}
        \caption{Mean flow.}
        \label{fig:base_meanaa}
    \end{subfigure}
    \hfill  
    \begin{subfigure}{0.46\textwidth}
        \centering
        \includegraphics[scale = 1]{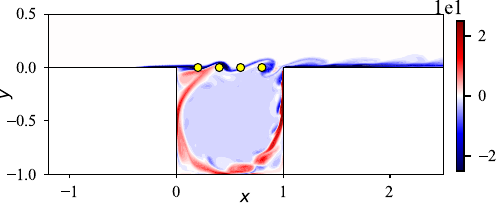}
        \caption {Vorticity snapshot.} 
        \label{fig:base_meanbb}
    \end{subfigure}
    \caption{Mean streamwise velocity field (a) and instantaneous vorticity field (b) for $Re=14000$.}
    \label{fig:base_mean_14000}
\end{figure}

\begin{figure}
    \centering
    \begin{subfigure}{0.337\textwidth}
        \centering
        \includegraphics[width=\linewidth]{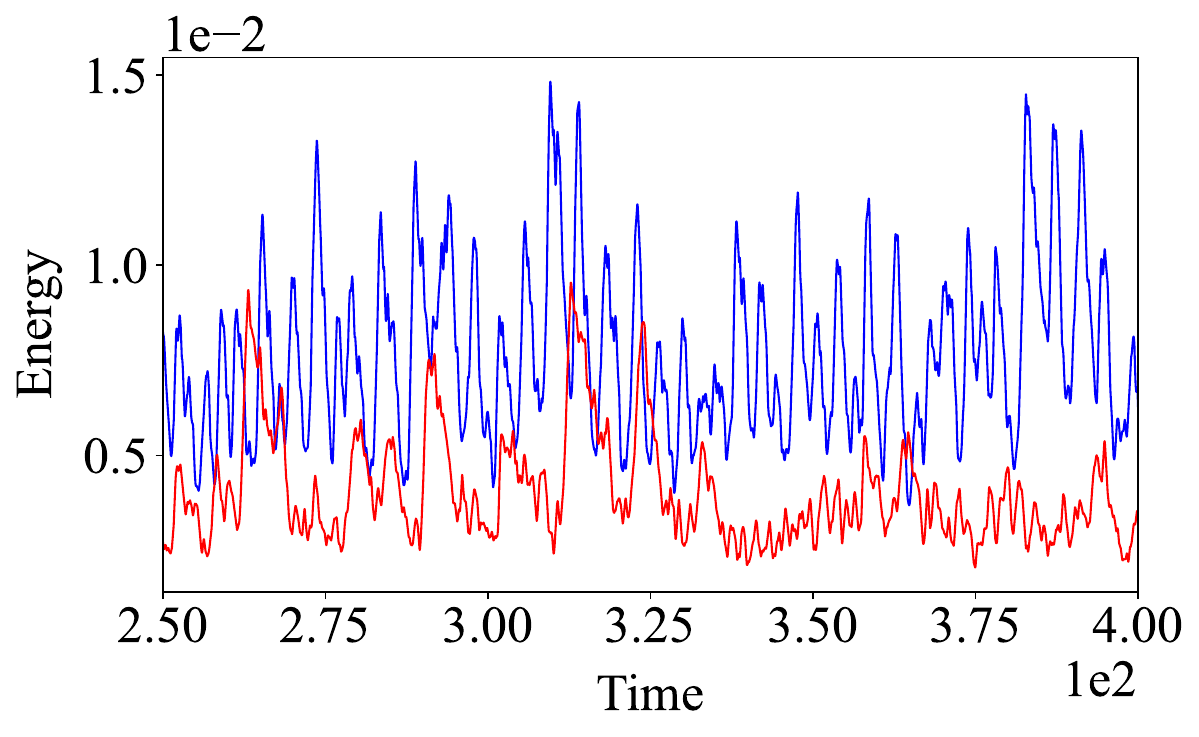}
        \caption{Time evolution of the global kinetic energy.}
        \label{fig:trajectory_shear_0_unc_14000a}
    \end{subfigure}
    \hfill
    \begin{subfigure}{0.337\textwidth}
        \centering
        \includegraphics[width=\linewidth]{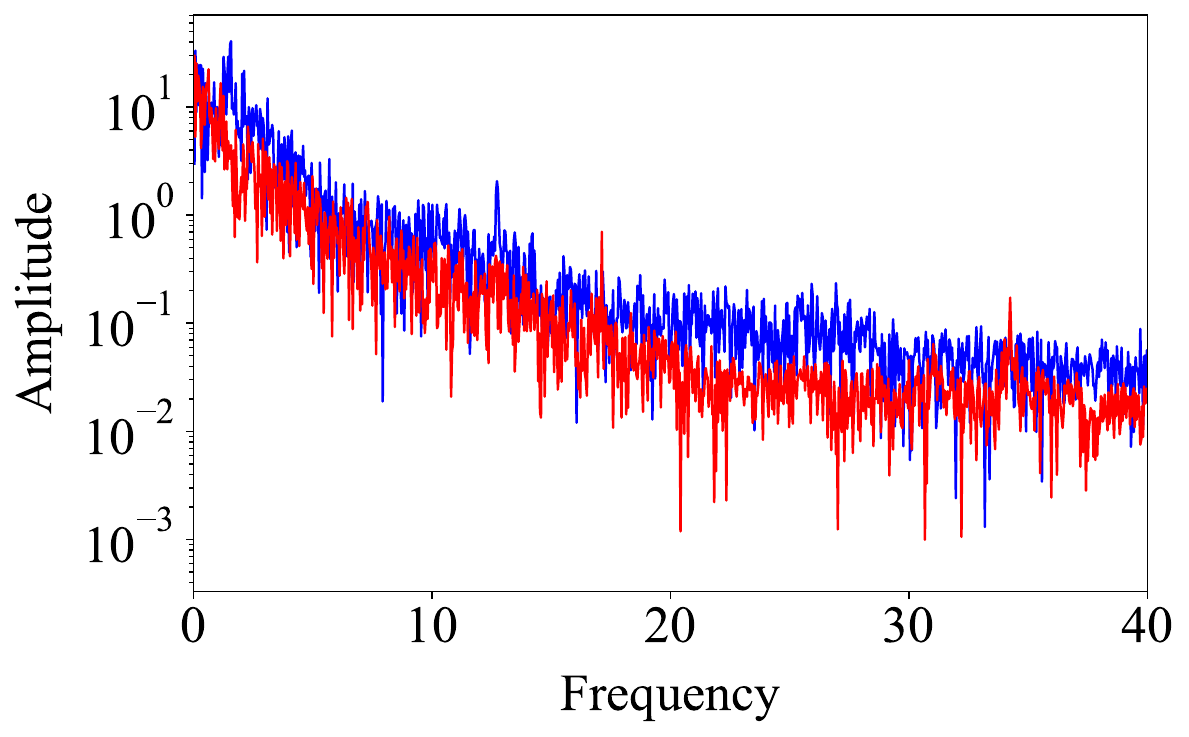}
        \caption{Frequency spectra.}
        \label{fig:trajectory_shear_0_unc_14000b}
    \end{subfigure}
    \hfill
    \begin{subfigure}{0.305\textwidth}
        \centering
        \includegraphics[width=\linewidth]{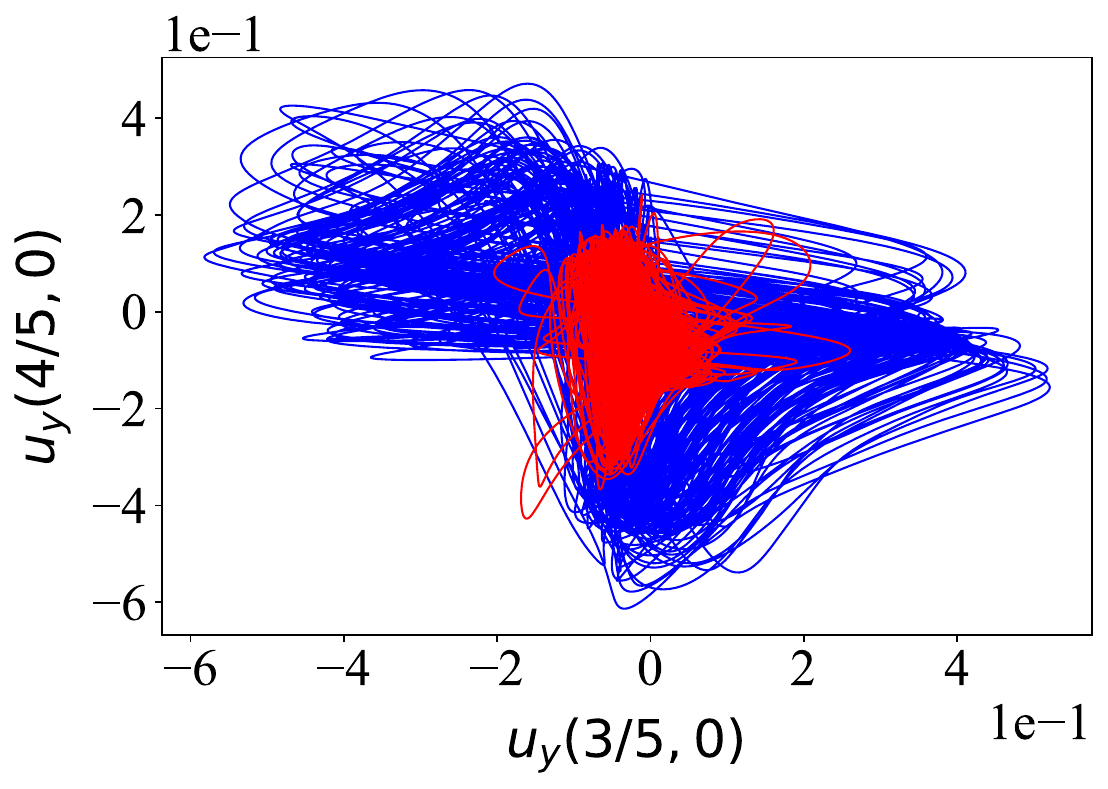}
        \caption{Trajectory in phase space.}
        \label{fig:trajectory_shear_0_unc_14000c}
    \end{subfigure}
    \caption{Characterisation of chaotic flow dynamics under controlled and uncontrolled conditions, at $Re=14000$. The uncontrolled flow is shown in blue, whereas the controlled flow obtained with EnVar L-BFGS method is shown in red. The time evolution of the global kinetic energy (a), the corresponding frequency spectra (b), and the trajectory constructed from $y$-velocity located in the shear layer at 
$y = 0$ and at streamwise positions $x = \tfrac{3}{5}$, $x = \tfrac{4}{5}$ (c) are reported.}
    \label{fig:trajectory_shear_0_unc_14000}
\end{figure}

Figure \ref{fig:base_meanaa} provides a first insight into the uncontrolled flow in this case. The mean streamwise velocity field highlights the main recirculation patterns inside the cavity and the development of the shear layer along the cavity opening. The vorticity snapshot illustrated in Figure \ref{fig:base_meanbb} suggests that, compared to the previous case $Re=6250$, the flow exhibits more pronounced oscillations and the shear layer destabilises earlier, reflecting the stronger unsteadiness at this higher Reynolds number. Flow fluctuations also appear to be more pronounced inside the cavity. These findings highlight the increased complexity of the flow dynamics at this Reynolds number, which is made more evident by the frequency spectra and phase-space trajectories presented in Figure \ref{fig:trajectory_shear_0_unc_14000}, as discussed in detail later in this section. 

\begin{figure}
    \centering
    \includegraphics[scale = 0.4]{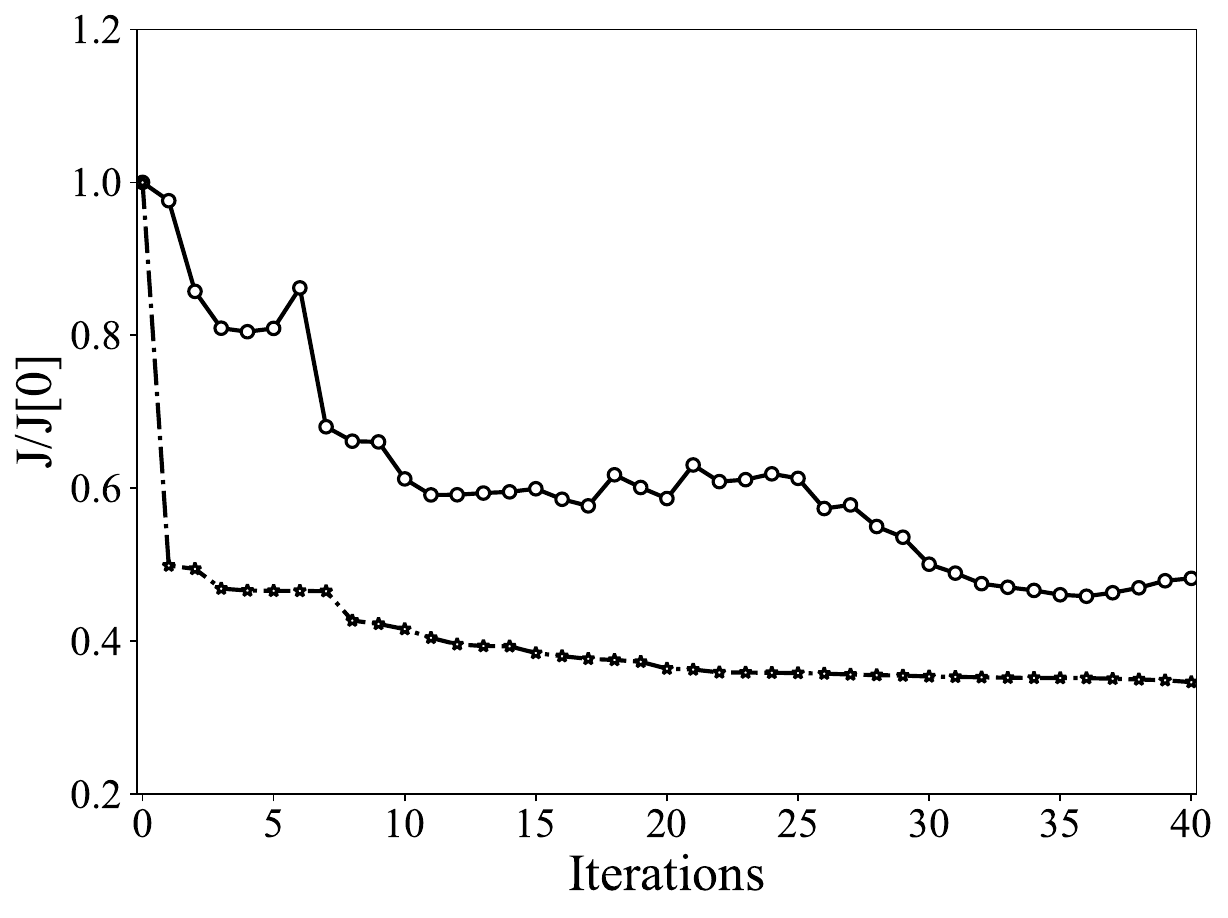}
    \caption{Evolution of the cost function for the EnVar Gauss--Newton method $-$ and the EnVar L-BFGS method $-\cdot-$, at $Re=14000$.}
    \label{fig:costfuncdecfreq014000}
\end{figure}

The setup of the control problem is the same as described in Section \ref{controlstrategy} and considered in the previous case. The cost function $J$ in \eqref{eq:specific_cost_function} is minimised with the EnVar method. Figure~\ref{fig:costfuncdecfreq014000} shows the evolution of the cost function throughout the EnVar-based optimisation performed with an ensemble size $N_{ens}=20$, obtained with both the Gauss--Newton and the L-BFGS algorithms. Compared to case $Re = 6250$, the Gauss--Newton approach requires a substantially larger number of iterations to achieve a meaningful reduction in the cost function, the evolution of the cost function being not strictly monotonic, with intermittent increases observed throughout the optimisation process. In contrast, the L-BFGS algorithm exhibits a significantly more robust behaviour. A pronounced reduction of the cost function is achieved already within the first iteration, followed by a rapid convergence towards an optimal solution. Moreover, the L-BFGS approach reaches a final value for the cost function that is significantly lower compared to the Gauss--Newton scheme. These better performances of the L-BFGS method compared to the Gauss-Newton one in the present regime might be attributed to the possibly more non-convex character of the cost function than in the case $Re=6250$, which is known to be better handled by quasi-Newton methods such as L-BFGS \cite{wang2017stochastic, cartis2010complexity}.

\begin{figure}
    \centering
    \includegraphics[scale = 0.9]{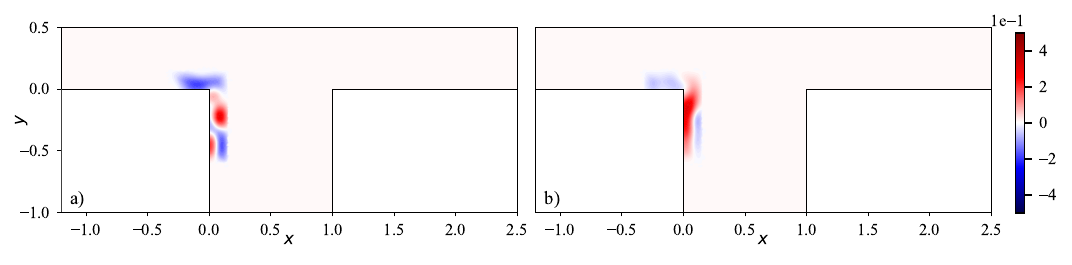}
    \caption{Optimal forcing obtained with EnVar Gauss--Newton (a) and with EnVar L-BFGS (b), at $Re=14000$. }
    \label{fig:ellipse1214000}
\end{figure}

The optimal forcing shapes obtained using the EnVar Gauss--Newton approach and the EnVar L-BFGS algorithm are compared in Figure \ref{fig:ellipse1214000}. The two solutions exhibit a number of common features; however, noticeable differences in the spatial distribution of the forcings are also observed, further confirming the existence of multiple locally optimal forcing distributions in the present regime. These differences also highlight the influence of the chosen optimisation algorithm on the resulting control shape, despite the use of the same ensemble-based gradient information.

The impact of the optimised steady forcing identified with the L-BFGS method (corresponding to the best optimisation results) on the flow dynamics is further investigated by comparing the controlled flow with the uncontrolled one in Figure \ref{fig:trajectory_shear_0_unc_14000}. Figure \ref{fig:trajectory_shear_0_unc_14000a} displays the time evolution of the global kinetic energy in \eqref{kinenergy}, highlighting a significant reduction in the flow fluctuations throughout the flow domain when control is applied. In Figure \ref{fig:trajectory_shear_0_unc_14000b}, the frequency spectra of the global kinetic energy signal are reported. The uncontrolled flow (in blue) exhibits broadband dynamics over a wide range of frequencies. The controlled flow (in red) shows a noticeable reduction in spectral amplitude (taking into account the logarithmic scale used in Figure \ref{fig:trajectory_shear_0_unc_14000b}) over most of the spectral range, confirming the efficacy of the actuation. The reduction in the amplitude of the flow fluctuations may also be assessed at a more local level through the phase portrait illustrated in Figure\ref{fig:trajectory_shear_0_unc_14000c} which is based on the same two transverse velocities in the shear layer as in the case $Re=6250$. From this figure, an attractor region might be identified for the uncontrolled case, whose extent is significantly decreased in the controlled one.


\begin{figure}
    \centering
    \includegraphics[scale = 0.9]{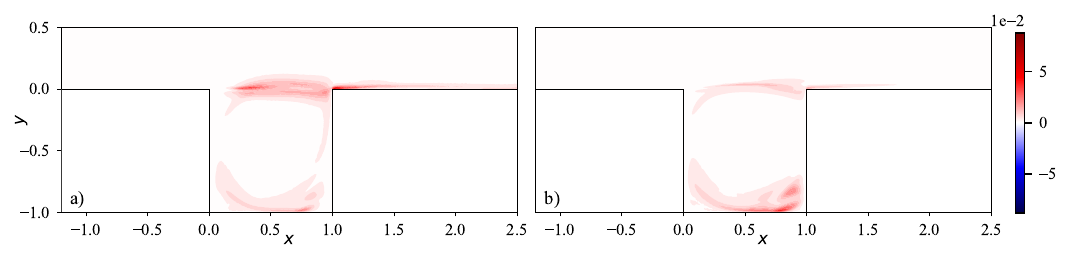}
    \caption{Reynolds stress component $\langle u'u' \rangle$ at $Re=14000$. Uncontrolled flow (a) and controlled flow obtained with EnVar L-BFGS method (b).}
    \label{STATS_uu_14000.pdf}
\end{figure}

\begin{figure}
    \centering
    \includegraphics[scale = 0.9]{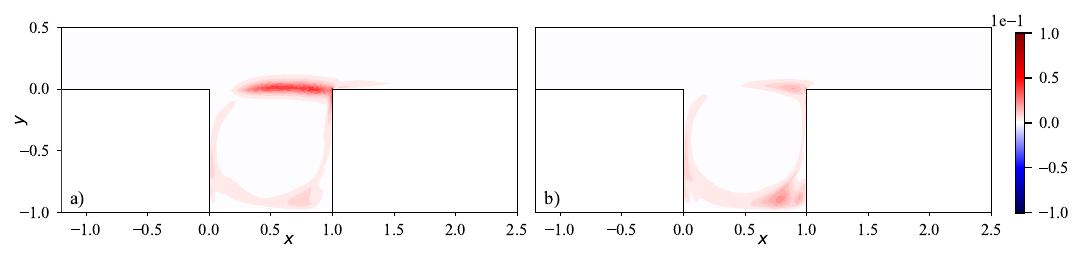}
    \caption{Reynolds stress component $\langle v'v' \rangle$ at $Re=14000$. Uncontrolled flow (a) and controlled flow obtained with EnVar L-BFGS method (b).}
    \label{STATS_vv_14000.pdf}
\end{figure}

Finally, the effect of the steady control on the statistical properties of the flow is examined. Figures \ref{STATS_uu_14000.pdf} and \ref{STATS_vv_14000.pdf} show the Reynolds stress components $\langle u'u' \rangle$ and $\langle v'v' \rangle$, respectively, comparing the uncontrolled flow with the controlled case. A strong amplitude reduction is observed for both components in the vicinity of the shear layer that extends over the opening of the cavity, especially for the Reynolds stress $\langle v'v' \rangle$. These observations are consistent with the spectral and trajectory analyses discussed above. However, it may be noticed that less intense fluctuations occurring at the bottom of the cavity have not been reduced (or even slightly increased) by the actuation. This possibly suggests the need to add velocity sensors in this region to ensure that such fluctuations are damped through the optimisation procedure, which is left for future work.

Overall, the results discussed in this subsection confirm the ability of the EnVar approach, especially when relying on a robust descent method, to deal with flows with complex dynamics and to identify forcing distributions that induce a significant damping of the flow fluctuations.

\subsection{On computational cost and parallelisation aspects}\label{sec:computational_cost}

In this subsection, we discuss the computational cost associated to the EnVar and adjoint approaches. Their respective costs can be expressed in terms of the number of direct/forward flow evaluations (integration of the Navier-Stokes equations over the considered time horizon) required. For the EnVar method, each iteration involves one direct simulation to evaluate the current estimate plus an additional $N_{ens}$ direct simulations to compute the ensemble perturbations. Thus, after $i$ iterations of the optimisation procedure, the overall cost scales as $(1+N_{ens})\,i$ direct-equivalent evaluations. In contrast, each iteration of the adjoint-based method requires one forward simulation and the corresponding backward adjoint solve. The cost of the latter depends on the implementation, in particular on the strategy used to access the forward trajectory during the backward integration. In the present adjoint solver, which relies on checkpointing (so that the computational cost of an adjoint calculation is roughly equivalent to that of $2$ forward ones), this leads to an approximate total cost of \(3\,i\) direct-equivalent evaluations. It should be noted that, in practice, the number of adjoint iterations can also depend on the specific optimisation algorithm used, which may include additional line-search or step-size evaluations.

An important distinction between the two approaches concerns their parallel efficiency. The ensemble members in the EnVar algorithm are independent and can be executed simultaneously on parallel hardware, reducing the wall-clock time of an iteration to nearly that of a single direct evaluation when sufficient computational resources are available. This property makes the method highly competitive, particularly for unsteady-flow optimisation problems. 
The adjoint method is less straightforward to parallelise, since the backward integration depends on the forward trajectory and often requires specific strategies for data storage and access. Parallel-in-time adjoint-based optimisation algorithms have been proposed \cite{Costanzo2022}, showing that, under suitable assumptions on memory availability and time partitioning, a significant reduction in computational cost can be achieved, up to $70\%$ of the total time of the optimisation process. The efficiency of such approaches, however, depends on the chosen time horizon, the communication overhead between processors, and the overall software implementation. In contrast, the EnVar approach relies on the independent propagation of ensemble members, which allows a direct exploitation of parallel computations and leads to a cost that scales more naturally with available resources.

Overall, when measured in terms of direct-equivalent cost, the adjoint method may require fewer simulations per iteration, but the intrinsic parallelism of EnVar can lead to substantially lower wall-clock times, making it an attractive alternative in high-performance computing environments.

\section{Conclusions and perspectives}\label{sectconcl}

In this work, an ensemble-variational (EnVar) approach has been proposed to extract steady flow-control strategies in an optimisation framework, focusing on high-dimensional parameter spaces that result from the spatial discretisation of an actuation field. A key aspect of the present approach is the use of a covariance matrix that is associated with the actuation, as justified by a Bayesian formulation of the flow-control problem, whose dominant eigenvectors are extracted to form the ensemble members in the EnVar framework. These eigenvectors enable the enforcement of spatial correlations and smoothness, while increasing the size of the ensemble may be interpreted as permitting the identification of smaller-scale features in the retrieved actuation. Sensitivities are evaluated in the ensemble subspace thus constructed, which is kept fixed among the iterations of the optimisation procedure. This avoids the need of regenerating ensemble members and undesirable effects such as ensemble collapse, thus favoring the robustness of the approach. Two descent methods are considered to exploit the computed sensitivities, namely the Gauss-Newton and L-BFGS algorithms.


This EnVar methodology has been assessed considering the flow past an open cavity with the aim of decreasing the intensity of flow fluctuations. Two regimes of increasing complexity have been examined: a quasi-periodic regime at $Re=6250$ and a chaotic one at $Re=14000$.
Results for the moderate Reynolds number regime ($Re=6250$) have confirmed that the EnVar approach is able to identify an optimal steady forcing distribution that is highly consistent with that obtained with a reference adjoint technique. A significant reduction in kinetic energy fluctuations has been achieved through the optimal forcing, effectively stabilizing the flow and driving its dynamics from a quasi-periodic state toward a periodic limit cycle. A sensitivity analysis has revealed that even with low ensemble sizes the dominant system sensitivities were captured, demonstrating the robustness of the EnVar approach with respect to this parameter.
The application to the chaotic regime at \(Re=14000\) has highlighted the critical role of the optimisation algorithm within the EnVar setting. In this more nonlinear regime, the cost function is expected to be more difficult to minimise, and the Gauss--Newton method has been shown to exhibit a slower reduction of the objective. By contrast, the L-BFGS algorithm has proven to be more robust, achieving rapid convergence and leading to a strong reduction of the Reynolds stresses, particularly near the cavity leading edge, confirming the ability of the EnVar approach to tackle complex flow dynamics.

From a computational perspective, the EnVar framework provides several notable advantages. Its non-intrusive formulation allows for direct coupling with existing high-fidelity CFD solvers, avoiding the development and maintenance of dedicated adjoint implementations. In addition, the ensemble evaluations are naturally parallelisable, leading to substantial reductions in wall-clock time relative to the inherently sequential forward–adjoint approach. These features make EnVar particularly well suited for large-scale engineering applications where computational resources are available but adjoint solvers are difficult to implement or suffer from numerical sensitivity. Moreover, the method remains applicable in complex physical regimes. 


In conclusion, this study demonstrates that EnVar methods provide a versatile and robust methodology for flow optimisation in complex unsteady regimes, even when considering high-dimensional parameter spaces. Future work will focus on extending this framework to three-dimensional configurations and investigating its performance in fully turbulent flows. Additionally, exploring time-dependent (harmonic) forcing strategies within the EnVar approach could further enhance the ability to control self-sustained oscillations and complex attractors across a broader range of fluid dynamics problems.

\section*{Acknowledgements}
The authors gratefully acknowledge the support provided by the Centre national d’études spatiales (CNES) (ROR: https://ror.org/04h1h0y33).




\appendix

\section{On the choice of the correlation-length coefficients}\label{appa}

In this appendix, we further discuss the choice of the correlation parameters involved in the expression of the prior covariance matrix $\mathbf{B}$ in \eqref{eq:covarianceB} ($a_x$, $a_y$ and $c_{xy}$). As a straightforward approach to evaluate the sensitivity of the EnVar approach with respect to these parameters, we evaluate the ability of the ensemble members, coinciding with the eigenvectors of \eqref{eq:covarianceB} (see Section \ref{sec:ensemblematrix}), to correctly represent a reference forcing $\boldsymbol{f}_{\rm ref}$ that is here chosen as the optimal solution identified by the adjoint approach in the case $Re=6250$ and illustrated in Figure \ref{fig:ellipse12}a. Namely, for a given ensemble size $N_{ens}$ and choice of the correlation parameters, we evaluate the following (relative) residual of the reference forcing with respect to the ensemble basis, or reconstruction error:
\begin{equation}\label{eq:reconstruction_error}
E_{\rm{rec}}=\frac{
\left( \boldsymbol{f}_{\rm ref} - \mathbf{U}_{N_{ens}} \mathbf{U}_{N_{ens}}^T \mathbf{M}_f \boldsymbol{f}_{\rm ref} \right)^{\mathrm{T}} 
\mathbf{M}_f 
\left( \boldsymbol{f}_{\rm ref} - \mathbf{U}_{N_{ens}} \mathbf{U}_{N_{ens}}^T \mathbf{M}_f \boldsymbol{f}_{\rm ref} \right)
}{
\boldsymbol{f}_{\rm ref}^{\mathrm{T}} \mathbf{M}_f \boldsymbol{f}_{\rm ref}
},
\end{equation}
where $\mathbf{U}_{N_{ens}}$ denotes the truncated matrix of eigenvectors of $\mathbf{B}$ corresponding to the $N_{ens}$ largest eigenvalues and $\mathbf{M}_f$ is associated to the discretisation of the scalar product for the forcing field (see Section \ref{sect3}). 

\begin{figure}
    \centering
    \includegraphics[scale = 0.5]{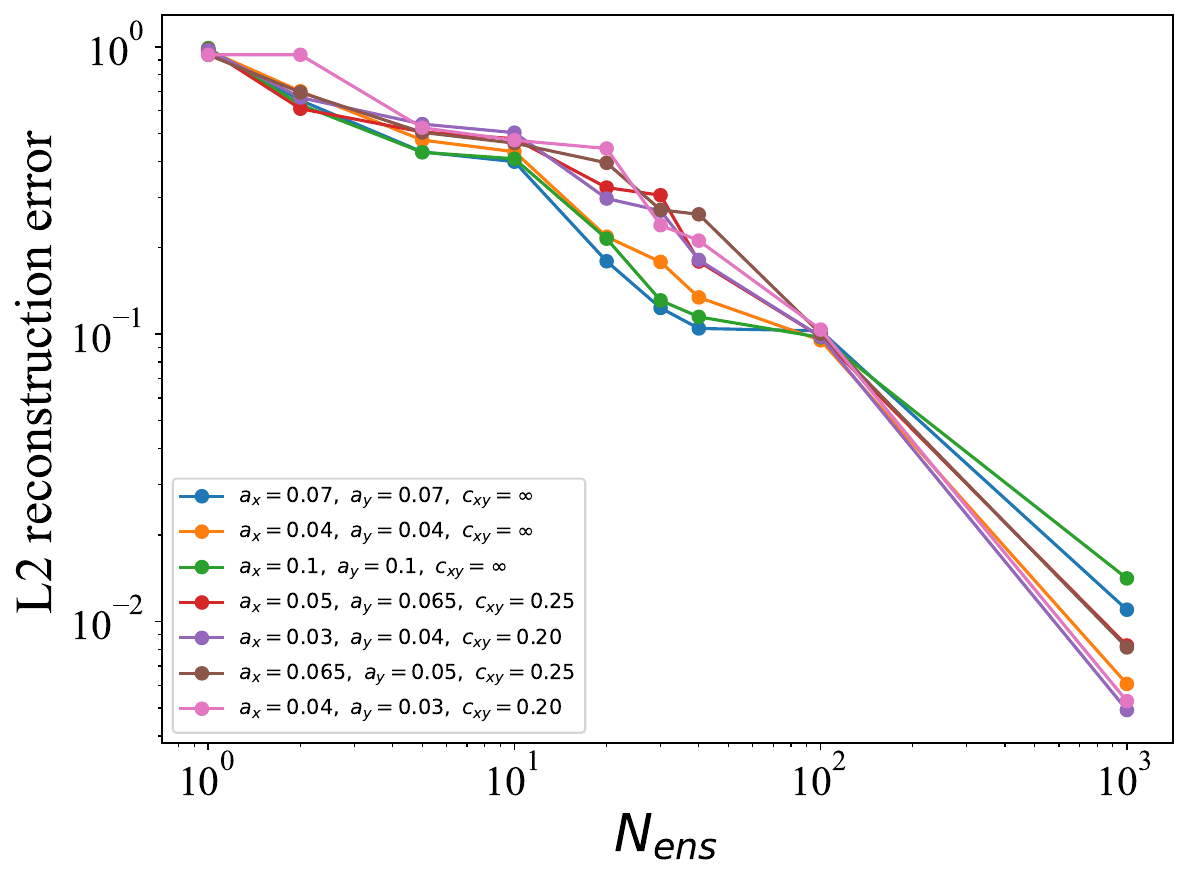}
    \caption{Reconstruction error in \eqref{eq:reconstruction_error} as a function of ensemble size $N_{ens}$ for different spatial correlation coefficients.}
    \label{fig:error01}
\end{figure}

Figure \ref{fig:error01} shows this normalized error as a function of the ensemble size $N_{ens}$ for various choices of spatial correlation coefficients. It appears that, for reasonable variations around the values chosen in Section \ref{sec:prior_parameters} ($a_x=a_y=0.07$ and $c_{xy}=\infty$, corresponding to the blue curve), the representativeness of the reference forcing in the ensemble basis is not significantly affected over an appreciable range of ensemble sizes. Accordingly, we do not expect the EnVar-based optimisation procedure to be noticeably sensitive with respect to these correlation parameters.


\section{Statistical analysis of the EnVar solutions}\label{appb}

We here further examine the optimal solutions obtained for the case \(Re=6250\) with the EnVar L-BFGS scheme for different ensemble sizes which are illustrated in Figure \ref{fig:ellipse10}. More specifically, we perform an uncertainty quantification study of these forcing solutions, as allowed by the Bayesian formalism, based on the associated posterior covariance matrix $\mathbf{B}_{post}$ given by \eqref{eq:posterior_covariance_matrix}.

\begin{figure}
    \centering
    \includegraphics[scale = 0.9]{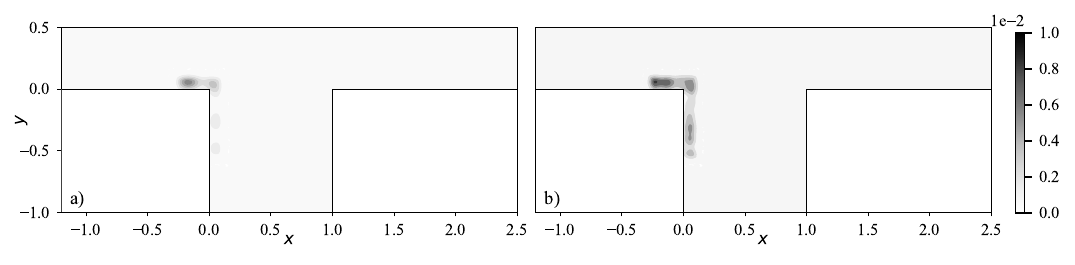}
        \includegraphics[scale = 0.9]{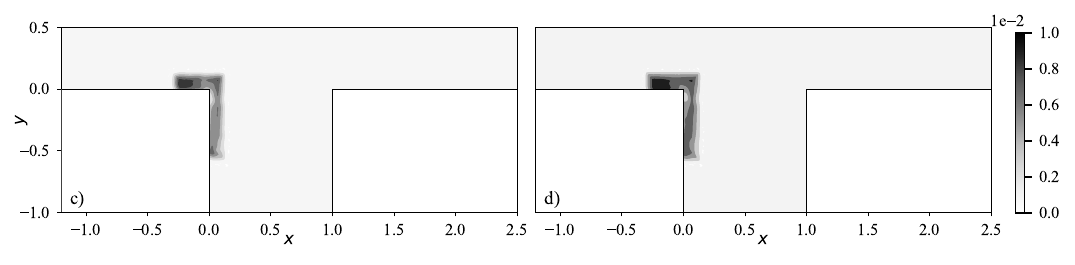}
    \caption{Estimated uncertainties obtained with EnVar L-BFGS method using 5 ensemble members (a),  10 ensemble members (b), 20 ensemble members (c),  40 ensemble members (d) for $Re=6250$.}
    \label{fig:ellipse12appb}
\end{figure}

The variance fields (extracted from the diagonal of $\mathbf{B}_{post}$) reported in Figure \ref{fig:ellipse12appb} exhibit a dependence on the ensemble size. As the number of ensemble members increases from \(N_{ens}=5\) to \(N_{ens}=40\), the estimated uncertainty becomes progressively richer and more spatially extended. This behaviour is expected, since a larger ensemble provides a more complete representation of the admissible perturbation subspace, whereas a small ensemble can only span a limited number of modes and therefore tends to underestimate the actual uncertainty level. 

For low values of \(N_{ens}\), the uncertainty field remains artificially confined, reflecting the limited ability of a reduced ensemble to capture the full variability permitted by the prior model. As \(N_{ens}\) increases, the uncertainty level increases in the regions that are less dynamically relevant to the optimisation and progressively approaches the prior distribution shown in figure \ref{fig:ellipse}. This suggests that, away from the most sensitive flow regions, the observations provide little constraint on the control, such that the posterior uncertainty remains close to the prior. 

By contrast, the region near the cavity edge consistently exhibits lower uncertainty, even for the largest ensemble sizes. This suggests that this region plays a dominant role in the optimisation process: the control there is more effectively constrained, and the solutions obtained with different ensemble sizes remain close to one another. The persistence of low uncertainty in this area reflects its dynamical importance, whereas the larger uncertainty observed elsewhere is associated with directions that are less influential on the objective functional.


\bibliography{references5}

\end{document}